\documentclass{jfm}
\usepackage{graphicx}
\usepackage{subfig}
\usepackage{epstopdf,epsfig}
\usepackage{newtxtext}
\usepackage{newtxmath}
\usepackage{natbib}
\usepackage{hyperref}
\usepackage{placeins}
\usepackage{enumitem}
\usepackage{cleveref}
\crefname{figure}{figure}{figures}
\crefname{equation}{}{}
\Crefname{equation}{Equation}{Equations}
\crefformat{section}{\S#2#1#3}
\crefmultiformat{section}{\S\S~#2#1#3}{ and~#2#1#3}{, #2#1#3}{, and~#2#1#3}

\hypersetup{
	bookmarksnumbered,
	colorlinks = true,
	linkcolor = blue,
	citecolor  = cyan
}
\graphicspath{{./Fig/},{./}}
\newcommand\bm{\boldsymbol}
\newcommand{\hatP}{\hat{P}}
\newcommand{\hatM}{\hat{M}}
\newcommand{\tildeP}{\tilde{P}}
\newcommand{\Wo}{\mbox{\textit{Wo}}}

\lefttitle{
	B. Wang et al.
	}
\righttitle{Journal of Fluid Mechanics}

\title{Taylor--Aris dispersion of active particles in oscillatory channel flows}

\author{
	Bohan Wang\aff{1},
	Weiquan Jiang\aff{2},
	Li Zeng\aff{1},
	Zi Wu\aff{3}
	\and
	Ping Wang\aff{4}
}

\affiliation{
	\aff{1} State Key Laboratory of Water Cycle and Water Security in River Basin, China Institute of Water Resources and Hydropower Research, Beijing 100038, China
	\aff{2}National Observation and Research Station of Coastal Ecological Environments in Macao; Macao Environmental Research Institute, Faculty of Innovation Engineering, Macau University of Science and Technology, Macao SAR 999078, China
	\aff{3} State Key Laboratory of Hydroscience and Engineering, Tsinghua University, Beijing 100084, China
	\aff{4} School of Soil and Water Conservation, Beijing Forestry University, Beijing 100083, China
}

\corresau{Li Zeng, \email{lizeng@iwhr.com}}

\begin{document}
\maketitle

\begin{abstract}
Mass dispersion in oscillatory flows is intimately linked to various environmental and biological processes, offering a distinct contrast to dispersion in steady flows due to the periodic expansion and contraction of particle patches. In this study, we investigate the Taylor--Aris dispersion of active particles in laminar oscillatory flows between parallel plates. Two complementary approaches are employed: a two-time-variable expansion of the Smoluchowski equation is used to facilitate Aris' method of moments for the preasymptotic dispersion, while the generalised Taylor dispersion theory is extended to capture phase-dependent periodic drift and dispersivity in the long-time asymptotic limit. Applying both frameworks, we find that spherical non-gyrotactic swimmers can exhibit greater or lesser diffusivity than passive solutes in purely oscillatory flows, depending on the oscillation frequency. This behaviour arise primarily from the disruption of cross-streamline migration governed by Jeffery orbits. When a steady component is superimposed, oscillation induces a non-monotonic dual effect on diffusivity. We further examine two well-studied shear-related accumulation mechanisms, arising from gyrotaxis and elongation. Although these accumulation effects are less pronounced than in steady flows due to flow unsteadiness, gyrotactic swimmers respond more effectively to the unsteady shear profile, significantly altering their drift and dispersivity. This work offers new insights into the dispersion of active particles in oscillatory flows and also provides a foundation for studying periodic active dispersion beyond the oscillatory flow, such as periodic variations in shape and swimming speed.
\end{abstract}


\section{Introduction}
Oscillatory flows are widely observed in various aquatic and biological systems, including tidal and wind-driven currents \citep{bars_flows_2015,wang_transport_2021}, tributary inflows influenced by diurnal discharge regulation \citep{long_tributary_2020}, biological flows \citep{secomb_blood_2017,lawrence_dispersion_2019,alaminos-quesada_effects_2024}, as well as in engineered systems \citep{reis_application_2006,vedel_pulsatile_2010,hacioglu_oscillating_2016}.
Compared with steady flows, mass transfer in oscillatory flows are more complex.
A patch of the substance can undergo expansion and contraction as the advective velocity profile develops and decays in opposite directions.
In this process, the extent to which the substance disperses along the streamwise direction depends on its cross-sectional passive diffusion or active migration, which governs its ability to utilise differential advection between streamlines.
Therefore, it is of interest to investigate how active particles, whose cross-sectional migration is jointly influenced by shear and motility \citep{bees_dispersion_2010,bearon_biased_2012,croze_dispersion_2013,croze_gyrotactic_2017}, disperse differently from solute particles, whose cross-sectional migration is purely diffusive and independent of the flow profile under laminar conditions.

Researches on dispersion of solute in oscillatory flows began shortly after the seminal work of \citet{taylor_dispersion_1953} on dispersion in steady pipe flow.
A significant milestone was achieved when \citet{aris_dispersion_1960} applied his method of concentration moments \citep{aris_dispersion_1956} to investigate the period-averaged asymptotic dispersion of solute in oscillatory pipe flow.
A key finding from relevant studies \citep{bowden_horizontal_1965,holley_dispersion_1970,fischer_mixing_1979} is that the effective period-averaged dispersivity in a purely oscillatory flow is significantly reduced compared to that in a steady flow with the same amplitude, due to concentration contraction when the flow direction reverses.
Building on \citet{aris_dispersion_1956}’s method of moments, \citet{brenner_macrotransport_1993} systematically developed the generalised Taylor dispersion (GTD) theory, a versatile framework capable of addressing both time- and space-dependent transport processes.
The core formulation of GTD involves the so-called \textit{a posterior} trial forms of the first- and second-order concentration moments, which, however, rely on intuitive judgement.

In contrast to the aforementioned studies on period-averaged asymptotic dispersion processes, \citet{chatwin_longitudinal_1975} focused on the variation within the oscillation period.
Through statistical analysis, he inferred that, as the solute tracers ultimately sample the entire cross-section, their time-dependent mean speed equals the instantaneous cross-sectional average of the streamwise advection velocity.
Based on this inference, he further conjectured that the time-dependent dispersivity oscillates at twice the frequency of the flow, due to the multiplicative interaction of time-dependent terms in the definition of mean square displacement.
Although \citet{chatwin_longitudinal_1975}'s formal proof of these conclusions relied on the simplifying assumption that the concentration varies linearly in the streamwise direction, his inferences were subsequently validated by other researchers \citep{yasuda_longitudinal_1984,mukherjee_dispersion_1988}.

Researchers have identified two key parameters governing the oscillatory dispersion of solute: the oscillatory period and the cross-sectional diffusion time scale.
Consider solute dispersion in an oscillatory flow driven by a time-periodic pressure gradient, with an initial uniform line source in the cross-section, released when the mean flow is zero.
During the first half of the oscillation cycle, the solute patch spreads in one streamwise direction.
If cross-sectional diffusion is relatively slow, the patch largely returns to its original position during the second half of the cycle, exhibiting only limited effective dispersion in the streamwise direction.
Therefore, transient negative dispersivity may arise in the later half of this expansion--contraction process, even though the period-averaged dispersivity remains positive.
This phenomenon motivated \citet{smith_contaminant_1982} to develop more a robust delayed-diffusion model to address the singularities caused by transient negative dispersivity.
\citet{yasuda_longitudinal_1984} later clarified the seemingly paradoxical transient negative dispersivity by proposing an alternative procedure for its calculation.
He compared the commonly used approach  --- averaging across the cross-section before calculating the mean square displacement --- with the newly proposed method, where the mean square displacement is calculated prior to averaging.
While the first method reproduces the previously observed transient negative dispersivity, the second method, which he regarded as more reasonable, consistently yields a positive transient dispersivity.

In light of the aforementioned research on dispersion in oscillatory flows, most studies assume simplified cross-sectional transport processes characterised by a constant and steady diffusivity. 
Under this setting, the cross-sectional distribution eventually becomes uniform, i.e., the zeroth-order longitudinal concentration moment (i.e., the marginal distribution of solute across the cross-section) approaches a uniform state, significantly simplifying the transient and asymptotic calculations.
For instance, based on an initial condition of uniform line release in the cross-section, \citet{ding_enhanced_2021} derived explicit expressions for the concentration moments up to the third order.
While this setting holds for solute dispersion in laminar oscillatory flows, it is not suitable for either active particles with non-diffusive migration \citep{fung_local_2022}, or for turbulent conditions with non-uniform cross-sectional diffusivity \citep{bowden_horizontal_1965}.

Beyond the special case of the ultimate uniform zeroth-order streamwise concentration moment in the cross-section, the GTD theory for time-periodic processes, proposed by \citet[Chapter 6]{brenner_macrotransport_1993}, addresses the dispersion of Brownian particles subjected to a time‐periodic external force in the cross-section \citep{haber_diffusion_1990,shapiro_taylor_1990,shapiro_taylor_1990a}.
In these circumstances, the zeroth-order streamwise concentration moment becomes time-periodic as well.
However, although the GTD theory for time-periodic processes of \citet{brenner_macrotransport_1993} has solved the time-dependent zeroth-order streamwise concentration moment, which adequately reflects the asymptotic transport mechanisms in the cross-section, it does not extend to how the drift and dispersivity behave asymptotically within the oscillation period, as only period-averaged solutions are obtained.

In this study, we consider the dispersion of active particles in an oscillatory channel flows between parallel plates driven by a time-periodic pressure gradient.
The active particles propel at a constant speed, referred to as swimmers hereafter, mimicking micro-organisms such as motile microalgae and bacteria.
The swimming directions of the swimmers are governed by the Jeffery equation \citep{jeffery_motion_1922}, with a possible modification from gyrotaxis, typically induced by bottom-heaviness of the swimmers \citep{pedley_hydrodynamic_1992}.
The original Jeffery equation describes the dependence of the angular velocity of a spheroid on the flow shear and rate-of-strain.
For elongated swimmers suspended in a steady laminar pressure-driven flow, the steady concentration distribution shows a non-trivial response to the mean shear rate \citep{vennamneni_shearinduced_2020}, as reported for both strongly elongated bacteria cells \citep{rusconi_bacterial_2014,bearon_trapping_2015} and less elongated algal cells \citep{barry_shear-induced_2015}.
On the other hand, gyrotactic swimmers also possess a shear-sensitive feature.
They are efficiently guided by the flow to accumulate near regions with the fastest downwelling speed \citep{kessler_cooperative_1985,kessler_hydrodynamic_1985}, which can lead to a modification on the flow profile and even hydrodynamic instabilities in relatively concentrated suspensions \citep{kessler_individual_1986,hwang_stability_2014,bees_advances_2020,fung_bifurcation_2020,fung_sequence_2020,ishikawa_instability_2022,fung_analogy_2023,wang_dispersion_2023,ishikawa_poiseuille_2025}.
Thus, both the elongated shape and gyrotaxis introduce a dependence of the swimmers' cross-sectional migration on the local time-periodic shear.
As a result, the existing transient solutions for the concentration moment equations \citep{mukherjee_dispersion_1988} cannot be directly applied due to the inherent shear-dependent transport in the orientation and position space.
Instead, a reformulation of the solutions for the moment equations, derived from the Smoluchowski equation for orientation--position coupled transport of swimmers in oscillatory flows, should be pursued to better understand the associated physics.
It is important to note that several orientational-averaged models have been developed to characterise swimmer transport of in position space, including the Pedley--Kessler model \citep{pedley_new_1990}, the two-step GTD model \citep{hill_taylor_2002,manela_generalized_2003}, and a more recent new model by \citet{fung_local_2022}.
However, these models are generally considered inapplicable when the flow shear varies rapidly, such as in high-frequency oscillatory flows \citep{caldag_fine-tuning_2025}.
Therefore, these models are not employed here.

For the long-time asymptotic dispersion regime, we revisit the classical GTD theory by \citep{brenner_macrotransport_1993}, with an extension to the asymptotic oscillatory behaviours of the drift and dispersivity during an oscillation period.
Such an extension is particularly useful for investigating the dispersion mechanism at a specific phase of the oscillation.
For the transient dispersion processes, we adopt the method of moments proposed by \citet{aris_dispersion_1956}.
It is important to note that the introduction of self-propulsion results in the non-self-adjointness of the eigenvalue problem, which is critical to the solution technique of \citet{barton_method_1983}.
This challenge is partially addressed by employing the biorthogonal expansion method \citep{strand_computation_1987,nambiar_stress_2019,jiang_transient_2021}.
However, the time-dependent eigenvalue problem becomes significantly more complex to solve, especially when an additional orientation space is involved.
Even for solute, related works addressing the time-dependent eigenvalue problem, or equally complex Green's function problem, have only derived the concentration moments up to the second order \citep{yasuda_longitudinal_1984,mukherjee_dispersion_1988,wu_environmental_2012}.
Therefore, while it is theoretically possible to solve for the transient dispersion of orientable swimmers in oscillatory flows, the dispersivity still presents significant technical challenges, let alone higher-order statistics.

Recently, \citet{jiang_transient_2025} applied a two-time-variable expansion to the concentration moment equations before solving them with eigenfunction expansion method \citep{barton_method_1983}, which is particularly efficient for higher-order concentration moments.
They introduced an auxiliary oscillation time variable that characterises the inherent oscillation in the dispersion due to the continuous expansion--contraction processes of the concentration.
The ideology underlying this new method actually shares a fundamental commonality with the GTD theory for asymptotic dispersion --- both feature an inherent periodic process governing the dispersion dynamics, while other transient modes gradually decay as the system evolves.
The present work extends the two-time-variable expansion method by \citet{jiang_transient_2025} to analyse the transient dispersion of swimmers in oscillatory flows, while also advancing the GTD theory to predict long-time asymptotic periodic behaviour of drift and dispersivity, rather than just their period-averaged values.

The remainder of this paper is organised as follows.
In \cref{sec:problem}, we describe the problem setup for swimmer dispersion in oscillatory flows between parallel plates.
In \cref{sec:transient solutions}, we develop the general solution procedure for the transient moment equations.
In \cref{sec:GTD}, we revisit the GTD theory for long-time asymptotic dispersion and extend it to account for phase-resolved drift and dispersivity.
In \cref{sec:results}, we present and discuss the results.
Finally, in \cref{sec:concluding remarks}, we provide some concluding remarks.

\section{Problem setup}
\label{sec:problem}
\subsection{Governing equation}
\begin{figure}
	\centerline{\includegraphics{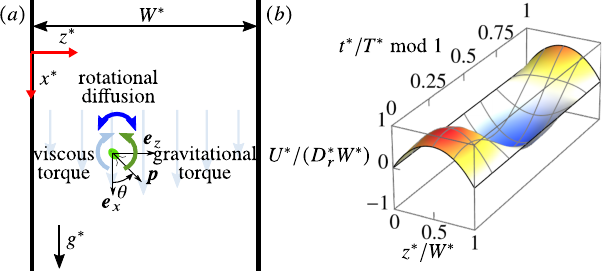}}
	\captionsetup{style=capcenter}
	\caption{
		Schematic illustration of swimmer dispersion in a vertical oscillatory channel flow.
		($a$): A gyrotactic swimmer in the oscillatory channel flow experiences a viscous torque and a gravitational torque, in addition to rotational diffusion.
		($b$): Time evolution of oscillatory velocity profile for the case with $P_0^{\ast} = 0$, $Q_0^{\ast} = 8 \nu^{\ast} D_r^{\ast}/W^{\ast}$, $\delta^{\ast} = 0.86 W^{\ast}$, and $\omega^{\ast} = D_r^{\ast}$.
	}
	\label{fig:sketch}
\end{figure}

As shown in \cref{fig:sketch}, we consider the dispersion of a patch of swimmers released into an oscillatory flow between two vertical, parallel plates separated at a distance of $W^{\ast}$.
Assuming a dilute suspension of swimmers, and thereby neglecting the forces exerted by the swimmer, the flow is driven by a pressure gradient composed of a steady part (which includes the contribution from gravity acceleration) and a zero-mean oscillatory part:
\begin{equation}
	G_p^{\ast}(z^{\ast}, t^{\ast}) = -P_0^{\ast} \bm{e}_x  - Q_0^{\ast} \cos \omega^{\ast} t^{\ast} \bm{e}_x.
\end{equation}
The velocity profile is given by \citep{vonkerczek_instability_1982}
\begin{gather}
	U^{\ast}(z^{\ast},t^{\ast}) = \frac{P_0^{\ast} (W^{\ast}-z^{\ast}) z^{\ast}}{2\nu^{\ast}} + \frac{Q_0^{\ast}}{\omega^{\ast}} \mathrm{Im}\left\{
	\left[
	1- \frac{ \cosh \left[\frac{(1+\mathrm{i}) (2z^{\ast}-W^{\ast})}{2 \delta^{\ast}}	\right]}{\cosh\left[\frac{(1+\mathrm{i}) W^{\ast}}{2\delta^{\ast}}\right] }
	\right]
	\mathrm{e}^{\mathrm{i} \omega^{\ast} t^{\ast}}
	\right\},
\end{gather}
where $\nu^{\ast}$ is the kinematic viscosity of water, $\omega^{\ast}$ is the angular frequency of oscillation, and $\delta^{\ast} = \sqrt{2\nu^{\ast}/\omega^{\ast}}$ denotes the thickness of the Stokes layer.

The swimmer propels itself at a constant speed $V_s^{\ast}$ along an unsteady direction described by a unit vector $\bm{p}$.
We note that the swimmers may exhibit more complex behaviours if they are allowed to rotate out of the $x$--$z$ plane, even when the flow velocity gradient exists solely in the $z$-direction.
An example is the resonate alignment of helical swimmers in oscillatory channel flows, as observed in \citet{hope_resonant_2016}.
As a first step in investigating oscillatory active dispersion, we therefore restrict our attention to particles whose rotation is constrained to the $x$--$z$ plane, for mathematical convenience.
In this case, the orientation of the swimmer is characterised by the angle $\theta$, defined as the counter-clockwise angle between $\bm{e}_x$ and $\bm{p}$, such that $\bm{p} = \cos\theta \bm{e}_x + \sin\theta \bm{e}_z$.

Apart from the random rotational diffusion, the deterministic rate of change of the swimmer's direction is influenced by the vorticity, rate-of-strain, and gravitational reorientation, as described by the modified Jeffery's orbit \citep{jeffery_motion_1922,pedley_hydrodynamic_1992}:
\begin{equation}
	\dot{\bm p}^{\ast} = \frac{1}{2} \bm{\omega}^{\ast} \times \bm{p} + \alpha_0 \bm{p} \bm{\cdot} \mathsfbi{E}^{\ast} \bm{\cdot} (\mathsfbi{I}- \bm{p\!p}) + \frac{1}{2B^{\ast}} \bm{p}\times \bm{k} \times \bm{p}.
\end{equation}
Here
\begin{equation}
	\bm{\omega^{\ast}} = \frac{\partial U^{\ast}}{\partial z^{\ast}} \bm{e}_z  \times \bm{e}_x
\end{equation}
denotes the vorticity,
$\alpha_0$ is the Bretherton parameter (with $\alpha_0=0$ for a sphere, and $\alpha_0=1$ for a slender rod),
\begin{equation}
	\mathsfbi{E}^{\ast} = \frac{1}{2} \frac{\partial U^{\ast}}{\partial z^{\ast}}\left( \bm{e}_x \bm{e}_z  + \bm{e}_z \bm{e}_z \right)
\end{equation}
is the rate-of-strain tensor, $\mathsfbi{I}$ is the identity tensor, $B^{\ast}$ is the gravitational reorientation time scale, and $\bm{k} = -\bm{e}_x$ is the unit vector directed opposite to gravity.

The governing equation of the probability density function (p.d.f.) $P$ of the swimmers takes the form of the Smoluchowski equation, with the dimensional form written as
\begin{gather}
	\frac{\partial P}{\partial t^{\ast}}+ \left(\frac{\partial}{\partial x^{\ast}} \bm{e}_x + \frac{\partial}{\partial z^{\ast}} \bm{e}_z \right) \bm{\cdot} \bm{J}_p^{\ast}+  \left(\frac{\partial}{\partial \theta} \bm{e}_{\theta} \right) \bm{\cdot} \bm{j}_r^{\ast}=0,
	\label{eq:transport equation dimensional}
\end{gather}
where
\begin{subequations}
\begin{equation}
	\bm{J}_p^{\ast} = [V_s^{\ast} \bm{p} + U^{\ast}(z^{\ast},t^{\ast}) \bm{e}_x] P- D_t^{\ast} \left(\frac{\partial P}{\partial x^{\ast}} \bm{e}_x + \frac{\partial P}{\partial z^{\ast}} \bm{e}_z \right),
\end{equation}
and
\begin{equation}
	\bm{j}_r^{\ast} = \dot{\bm p}^{\ast} P-D_r^{\ast}  \frac{\partial P}{\partial \theta} \bm{e}_{\theta}
\end{equation}
\end{subequations}
are the fluxes in the position and orientation spaces, respectively.
Here $D_r^{\ast}$ and $D_t^{\ast}$ represent the rotational and translational diffusivity, respectively.

We non-dimensionalise the problem by
\begin{align}
	\left.
	\begin{array}{c}
		\displaystyle
		t=t^{\ast}D_r^{\ast},\quad \omega = \frac{\omega^{\ast}}{D_r^{\ast}}, \quad z = \frac{z^{\ast}}{W^{\ast}}, \quad x=\frac{x^{\ast}}{W^{\ast}},\quad \Wo =  \frac{W^{\ast}}{\delta^{\ast}}, \quad U=\frac{U^{\ast}}{D_r^{\ast} W^{\ast}},\\[12pt]
		\displaystyle
		\Pen_s=\frac{V_s^{\ast}}{D_r^{\ast}W^{\ast}},\quad \Pen_f^{s} = \frac{U_s^{\ast}}{D_r^{\ast}W^{\ast}},\quad \Pen_f^{o} = \frac{U_o^{\ast}}{D_r^{\ast}W^{\ast}}, \quad \lambda = \frac{1}{2B^{\ast} D_r^{\ast}} ,\quad D_t=\frac{ D_t^{\ast}}{D_r^{\ast}{W^{\ast}}^2}.
	\end{array}
	\right\}
	\label{eq:nondimensionalisation}
\end{align}
Here, the characteristic velocities $U_s^{\ast} = {W^{\ast}}^2 P_0^{\ast}/(8\nu^{\ast})$ and $U_o^{\ast} = {W^{\ast}}^2 Q_0^{\ast}/(8\nu^{\ast})$ are defined based on the steady and oscillatory pressure gradients, respectively.
In \cref{eq:nondimensionalisation}, $t$ denotes the dimensionless time, rescaled with the rotational diffusion time scale; $\omega$ is the angular frequency of oscillation non-dimensionalised by the rotational diffusivity; $z$ and $x$ are the cross-sectional and streamwise coordinates, respectively, rescaled with the channel width; $\Wo$ is the Womersley number; $U$ is the dimensionless flow velocity rescaled by the characteristic velocity associated with traversing the channel width within one rotational diffusion time scale; $\Pen_s$ is the dimensionless swimming P\'eclet number; $\Pen_f^s$ and $\Pen_f^o$ correspond to the steady and oscillatory flow P\'eclet numbers, respectively; $D_t$ is the dimensionless translational diffusivity rescaled with the characteristic diffusivity associated with diffusing across the channel width within one rotational diffusion time scale; and $\lambda$ is the dimensionless gravitactic bias parameter.

The dimensionless Smoluchowski equation is given by
\begin{gather}
	\frac{\partial P}{\partial t} + (U + \Pen_s \cos\theta) \frac{\partial P}{\partial x} - D_t \frac{\partial^2 P}{\partial x^2} + \Pen_s \sin\theta \frac{\partial P}{\partial z} - D_t \frac{\partial^2 P}{\partial z^2} + \frac{\partial (\dot{\theta} P)}{\partial \theta} - \frac{\partial^2 P}{\partial \theta^2} = 0, 
	\label{eq:transport equation dimensionless}
\end{gather}
where the dimensionless rate of change of $\theta$ is
\begin{gather}
	\dot{\theta}(z,\theta,t) = \frac{1}{2} \frac{\partial U}{\partial z} (-1 + \alpha_0 \cos 2 \theta) + \lambda \sin\theta,
\end{gather}
and the dimensionless flow velocity is
\begin{gather}
	U(z,t) = 4\Pen_f^{s}(1-z)z +  \frac{4 \Pen_f^{o}}{\Wo^2}  \mathrm{Im}\left\{
	\left[
	1- \frac{\cosh\left[\Wo(1+\mathrm{i}) (2z-1)/2\right]}{\cosh[\Wo (1+\mathrm{i})/2]}
	\right]
	\mathrm{e}^{\mathrm{i} \omega t}
	\right\}.
\end{gather}
We further define $T = 2\upi/\omega$ as the dimensionless oscillation period.

\subsection{Periodic, boundary and initial conditions}
The periodic conditions in the orientation variable $\theta$ are naturally satisfied:
\begin{subequations}
\begin{equation}
	P|_{\theta=0} = P|_{\theta=2\upi},
\end{equation}
\begin{equation}
	\frac{\partial P}{\partial \theta}|_{\theta=0}= \frac{\partial P}{\partial \theta}|_{\theta=2\upi}.
\end{equation}
\end{subequations}

The interaction between swimmer and wall is inherently complex \citep{maretvadakethope_interplay_2023}, involving potential wall-induced modifications to swimming speed, angular velocity, and rotational diffusivity \citep{kantsler_ciliary_2013,zeng_sharp_2022}.
However, since the primary aim of this work is to elucidate the dispersion mechanism governed predominantly by swimming and oscillatory shear, we adopt one of the most idealised and widely used boundary conditions in the continuum models --- the reflective boundary conditions:
\begin{subequations}
\begin{equation}
	P|_{\theta=\theta_0} = P|_{\theta=2\upi-\theta_0},\quad \mbox{on} \ z=0, 1,
\end{equation}
\label{eq:reflective boundary conditions P}
\begin{equation}
	\frac{\partial P}{\partial z}|_{\theta=\theta_0}= - \frac{\partial P}{\partial z}|_{\theta=2\upi-\theta_0},\quad \mbox{on} \ z=0, 1.
\end{equation}
\label{eq:reflective boundary conditions pPpz}
\end{subequations}
The reflective boundary conditions ensure the no-flux condition at the channel walls:
\begin{equation}
	\int_0^{2\upi} \left( \Pen_s \sin\theta \frac{\partial P}{\partial z} - D_t \frac{\partial^2 P}{\partial z^2} \right) \,\mathrm{d} \theta = 0,\quad \mbox{on} \ z=0, 1,
\end{equation}
thereby guaranteeing the conservation of the total number of swimmers.

The initial condition for $P(x,z,\theta,t)$ is formally prescribed as a source located in $x=0$:
\begin{equation}
	P|_{t=0} = I_{ini}(z, \theta) \delta(x), 
	\label{eq:initial condition for P}
\end{equation}
where $I_{ini}(z,\theta)$ represents the initial distribution in the cross-section-orientation space and satisfies the normalisation condition
\begin{equation}
	\int_0^1 \int_0^{2 \upi} I_{ini}(z, \theta) \, \mathrm{d}\theta \mathrm{d} z = 1.
\end{equation}

\section{Two-time-variable expansion for the transient moments}
\label{sec:transient solutions}
\subsection{Smoluchowski equation in the two-time-variable system}
Following \citet{jiang_transient_2025}, we introduce a two-time-variable expansion into the problem --- the basic time variable and an auxiliary oscillatory time variable:
\begin{equation}
	t_0 \triangleq t, \quad t_1 \triangleq \omega t.
	\label{eq:two time variables}
\end{equation}
The auxiliary oscillatory time variable $t_1$ is introduced to characterise the intrinsic periodic oscillation of the flow and is applied exclusively to the time-periodic oscillatory advection term, thus the advection velocity becomes
\begin{equation}
	U(z,t_1) =4\Pen_f^{s}(1-z)z +  \frac{4 \Pen_f^{o}}{\Wo^2}  \mathrm{Im}\left\{
	\left[
	1- \frac{\cosh\left[\Wo(1+\mathrm{i}) (2z-1)/2	\right]}{\cosh[\Wo (1+\mathrm{i})/2]}
	\right]
	\mathrm{e}^{\mathrm{i} t_1}
	\right\}.
\end{equation}

As the original time variable $t$ splits into two new time variables, $t_0$ and $t_1$, the p.d.f. must be redefined accordingly to reflect its dependence on both time scales:
\begin{equation}
	\hatP(x,z,\theta,t_0,t_1) \triangleq P(x,z,\theta,t=t_0),
	\label{eq:equivalence between P hatP}
\end{equation}
where the hat notation is used to distinguish the new two-time-variable p.d.f. from the original p.d.f..

Since $t_1$ is introduced to characterise the time-periodic behaviour of the system, we impose periodicity of $\hatP$ in $t_1$ as
\begin{equation}
	\hatP(x,z,\theta,t_0,t_1) =  \hatP(x,z,\theta,t_0,t_1+2\pi).
\end{equation}
Accordingly, we map $t_1$ to the interval $[0,2\upi)$ via
\begin{equation}
	t_1 = t_1  \bmod 2\upi.
\end{equation}

The use of two time variables offers two main advantages.
First, it circumvents the complexity associated with solving the time-dependent eigenvalue problem, such as in \citet{mukherjee_dispersion_1988}.
Second, it clarifies the underlying physical interpretation of the dispersion problem: $t_0$ represents the time scale associated with the evolving dispersion from the initial release, while $t_1$ captures the time-periodic dispersion process due to the flow oscillation.
As will be further discussed in \cref{sec:GTD}, this approach aligns naturally with the GTD theory, wherein $t_1$ emerges as the sole relevant time variable governing the asymptotic dispersion regime.

Under the two-time-variable expansion, the time derivative transforms into
\begin{equation}
	\frac{\partial}{\partial t} \to \frac{\partial}{\partial t_0} + \omega \frac{\partial}{\partial t_1}.
	\label{eq:split of time derivative}
\end{equation}
It is worth noting that the splitting of time derivative has also been employed in several studies investigating the emergent dynamics of single swimmers with periodically varying shape and/or speed \citep{gaffney_canonical_2022,walker_effects_2022,walker_emergent_2022,walker_systematic_2023,dalwadi_generalised_2024,dalwadi_generalised_2024a}, with a most recent work further considering a periodic flow field \citep{gaffney_motility_2025}.
However, the methodologies diverge from this point on.
In the multiple-time-scale approach, the periodic variation in shape or speed is assumed to be fast ($\omega \gg 1$), which naturally defines $t_1$ as a fast time scale to facilitate perturbation analysis.
In contrast, the two-time-variable method employed here does not assume any separation of time scale between $t_0$ and $t_1$.

Another less directly related category of works focuses on deriving effective evolution equations for the concentration of swimmers populations subject to rotational and/or translational noises.
Representative works include \citet{bearon_trapping_2015}, \citet{vennamneni_shearinduced_2020}, and \citet{fung_local_2022}, all of which also employ multi-time-scale perturbation analysis.
These studies typically introduce a slow time scale to derive effective transport equations tailored to different physical scenarios. 
However,  none explicitly consider time-periodic transport processes. 
Instead, their scale hierarchy is based on the small swimming P\'eclet number, which compares the mean straight swimming length to the confinement scale.

Substituting \cref{eq:split of time derivative} into the original transport equation \cref{eq:transport equation dimensionless} yields the two-time-variable formulation of the Smoluchowski equation:
\begin{gather}
	\frac{\partial \hatP}{\partial t_0} + \omega \frac{\partial \hatP}{\partial t_1}+ U_x \frac{\partial \hatP}{\partial x} - D_t \frac{\partial^2 \hatP}{\partial x^2} + \Pen_s \sin\theta \frac{\partial \hatP}{\partial z} - D_t \frac{\partial^2 \hatP}{\partial z^2} + \frac{\partial (\dot{\theta} \hatP)}{\partial \theta} - \frac{\partial^2 \hatP}{\partial \theta^2} = 0, \label{eq:expanded equation}
\end{gather}
where
\begin{equation}
	U_x(z,\theta,t_1) = U(z,t_1) + \Pen_s \cos\theta
\end{equation}
is the streamwise translational velocity, comprising both oscillatory advection and autonomous swimming.

It is important to note that, by implementing the two-time-variable expansion and imposing periodic conditions on the oscillatory time variable $t_1$, this variable can effectively be regraded and treated as a pseudo-spatial variable defined on the interval $[0,2\upi)$, associated with a convection term $\omega (\partial (\cdot)/\partial t_1)$.
The absence of a diffusion term in \cref{eq:expanded equation} for $t_1$ actually is a direct consequence of intrinsic relationship between $t_1$ and $t_0$: specifically, $t_1$ evolves proportional with $t_0$ at a rate governed by $\omega$, which manifests solely through the convective term in the transport equation.
We also note that the definition of $t_0$ relates bidirectionally with the original time variable $t$, whereas $t_1$ relates unidirectional from $t$.
Consequently, specifying $t_0$ uniquely determines the original single time variable $t$, while specifying $t_1$ alone does not.

In summary, the application of the two-time-variable expansion appears to increase the dimensionality of the Smoluchowski equation by one.
However, this transformation facilitates the subsequent solution of moment equations.
\Cref{eq:expanded equation} also underscores a key distinction between passive and active dispersion: swimmer motility causes the streamwise translation velocity and cross-sectional migration velocity to depend on not only the position but also on the orientation.
Moreover, orientation itself is influenced by these velocities, resulting in a bidirectional coupling.

\subsection{Moment equations in the two-time-variable system}
The $n$th-order local p.d.f. moment is defined as
\begin{equation}
	\hatP_n(z,\theta,t_0,t_1) \triangleq \int_{-\infty}^{\infty} x^n \hatP \,\mathrm{d}x.
\end{equation}
The governing equations for the first three local p.d.f. moments, $\hatP_0$, $\hatP_1$, and $\hatP_2$ can be derived from \cref{eq:expanded equation}, under the assumptions that $\hatP \to 0$ and $\partial \hatP/\partial x \to 0$ as $|x| \to \infty$:
\begin{subequations}
\begin{equation}
	\frac{\partial \hatP_0}{\partial t_0} + \omega \frac{\partial \hatP_0}{\partial t_1}  + \mathcal{L}_{co} \hatP_0 = 0,\label{eq:governing P0}
\end{equation}
\begin{equation}
	\frac{\partial \hatP_1}{\partial t_0} + \omega \frac{\partial \hatP_1}{\partial t_1}+\mathcal{L}_{co} \hatP_1 = U_x \hatP_0,\label{eq:governing P1}
\end{equation}
\begin{equation}
	\frac{\partial \hatP_2}{\partial t_0} + \omega \frac{\partial \hatP_2}{\partial t_1}+\mathcal{L}_{co} \hatP_2 =  2D_t\hatP_0 +  2U_x \hatP_1,\label{eq:governing P2}
\end{equation}
\end{subequations}
where $\mathcal{L}_{co}$ denotes the flux operator in the cross-section-orientation $(z,\theta)$ space:
\begin{equation}
	\mathcal{L}_{co}(\cdot) \triangleq \Pen_s \sin\theta \frac{\partial (\cdot)}{\partial z} - D_t \frac{\partial^2 (\cdot)}{\partial z^2} + \frac{\partial({\dot{\theta} (\cdot)})}{\partial \theta} - \frac{\partial^2 (\cdot)}{\partial \theta^2}.
\end{equation}

The periodic conditions satisfied by $P$ in $\theta$-space are inherited by $\hatP_n$:
\begin{subequations}
\begin{equation}
	\hatP_n|_{\theta=0} = \hatP_n|_{\theta=2\upi},
\end{equation}
\begin{equation}
	\frac{\partial \hatP_n}{\partial \theta}|_{\theta=0}= \frac{\partial \hatP_n}{\partial \theta}|_{\theta=2\upi}.
\end{equation}
\end{subequations}
Similarly, $\hatP_n$ satisfies the periodicity in $t_1$:
\begin{equation}
	\hatP_n|_{t_1=0} = \hatP_n|_{t_1=2\upi}.
\end{equation}

The reflective boundary conditions imposed on $P$ in the $(z,\theta)$ space also apply to $\hatP_n$:
\begin{subequations}
\begin{equation}
	\hatP_n|_{\theta=\theta_0} = \hatP_n|_{\theta=2\upi-\theta_0},\quad \mbox{on} \ z=0, 1,
\end{equation}
\begin{equation}
	\frac{\partial \hatP_n}{\partial z}|_{\theta=\theta_0} = - \frac{\partial \hatP_n}{\partial z}|_{\theta=2\upi-\theta_0},\quad \mbox{on} \ z=0, 1.
\end{equation}
\end{subequations}
	
For the initial condition of $\hatP$, the following relation can be deduced based on \cref{eq:equivalence between P hatP,eq:two time variables}:
\begin{equation}
	\hatP|_{t_0=0,t_1=0} = P|_{t=0}.
	\label{eq:initial condition requirement}
\end{equation}
Thus, we use
\begin{equation}
	\hatP|_{t_0=0} = I_{ini}(z,\theta) \delta(x)
\end{equation}
to satisfy \cref{eq:initial condition requirement}, with the normalisation condition:
\begin{equation}
	\frac{1}{2\upi} \int_{-\infty}^{\infty} \int_{0}^{1} \int_{0}^{2\upi} \int_{0}^{2\upi} \hatP|_{t_0=0} \,\mathrm{d}t_1\, \mathrm{d}\theta\, \mathrm{d}z\,\mathrm{d}x = 1.
	\label{eq:normalisation P}
\end{equation}
The initial conditions for the moments are:
\begin{subequations}
\begin{equation}
	\hatP_0|_{t_0=0} = I_{ini}(z,\theta),\label{eq:initial condition P0}
\end{equation}
\begin{equation}
	\hatP_1|_{t_0=0} = 0,
\end{equation}
\begin{equation}
	\hatP_2|_{t_0=0} = 0.
\end{equation}
\end{subequations}

The total moments integrated over $(z,\theta)$ are defined as
\begin{gather}
	\left\langle \hatP_n\right\rangle_{z,\theta}(t_0,t_1)= \int_0^1 \int_0^{2\upi} \hatP_n \,\mathrm{d}\theta \,\mathrm{d}z,
\end{gather}
with the corresponding governing equations:
\begin{equation}
	\frac{\partial \left\langle \hatP_n\right\rangle_{z,\theta}}{\partial t_0} + \omega \frac{\partial \left\langle \hatP_n\right\rangle_{z,\theta}}{\partial t_1} = n(n-1) D_t \left\langle \hatP_{n-2} \right\rangle_{z,\theta} + n \left\langle U_x \hatP_{n-1}  \right\rangle_{z,\theta}.
	\label{eq:governing equations Pn integrated over z theta}
\end{equation}
Note that we have implicitly used the relations that $\left\langle \mathcal{L}_{co} \hatP_n \right\rangle_{z,\theta} = 0$, based on the no-flux condition at the boundary and the periodicity in $\theta$.
It is noted that $\left\langle \hatP_n\right\rangle_{z,\theta}$ can be interpreted as the total moments in the single-time-variable dispersion problem, which are also used to calculate the transient evolution of drift and dispersivity.

A further integration of \cref{eq:governing equations Pn integrated over z theta} over $t_1 \in [0,2\upi]$, followed by division by $2\upi$, yields
\begin{equation}
	\frac{\partial\hatM_n}{\partial t_0}  = n(n-1) D_t \hatM_{n-2} + n  \overline{\left\langle U_x \hatP_{n-1} \right\rangle_{z,\theta}},
	\label{eq:governing equations Mn}
\end{equation}
where
\begin{equation}
	\overline{(\cdot)} \triangleq \frac{1}{2\upi}  \int_0^{2\upi}   (\cdot) \,\mathrm{d}t_1
\end{equation}
represents the mean operation over $t_1$, and
\begin{equation}
	\hatM_n = \overline{\left\langle \hatP_n \right\rangle_{z,\theta}} \label{eq:Mn Pn}
\end{equation}
can be interpreted as the total moments in the two-time-variable system viewing $t_1$ as a periodic variable.
\Cref{eq:governing equations Mn} describes how the moments evolve from a period-averaged perspective.

\subsection{Eigenfunction expansions and biorthogonality relation}
Up to this point, we can solve the equations \cref{eq:governing P0,eq:governing P1,eq:governing P2} successively, subject to the appropriate boundary and initial conditions.
Next, we define the new effective flux operator $\mathcal{L}_{coo}$ in the cross-section-orientation-oscillation space ($z,\theta,t_1$), incorporating the advection term in $t_1$:
\begin{equation}
	\mathcal{L}_{coo}(\cdot) \triangleq \mathcal{L}_{co}(\cdot) + \omega \frac{\partial(\cdot)}{\partial t_1}.
\end{equation}
To solve for the moment equations using \citet{barton_method_1983}'s eigenfunction expansion method, we seek to solve the eigenvalue problem:
\begin{equation}
	\mathcal{L}_{coo} f_i  = \lambda_i f_i, \label{eq:eigenvalue problem}
\end{equation}
where $f_i$ are the eigenfunctions and $\lambda_i$ are the corresponding eigenvalues.
Due to the orientable motility of swimmers, $\mathcal{L}_{coo}$ is non-self-adjoint; therefore, we employ a Galerkin method to numerically solve \cref{eq:eigenvalue problem}.
The main steps involved in this approach are as follows:
\begin{enumerate}[leftmargin=2.5em, labelsep=0.5em, label=(\roman*)]
	\item Finding the appropriate basis functions in the $(z,\theta,t_1)$ space that satisfy the reflective boundary conditions \cref{eq:reflective boundary conditions P,eq:reflective boundary conditions pPpz} in $(z,\theta)$ space, and the periodicity in $t_1$ space.
	These basis functions are denoted as $\{e_i\}_{i=1}^{\infty}$, with the normalisation condition:
	\begin{equation}
		\int_{0}^{1} \int_{0}^{2\upi} \int_{0}^{2\upi} e_i e_j \,\mathrm{d}t_1\, \mathrm{d}\theta\, \mathrm{d}z = \delta_{ij},
	\end{equation}
	where $\delta_{ij}$ is the Kronecker delta function.
	The basis functions satisfying the reflective boundary conditions, as given in \citet{wang_gyrotactic_2022} for steady channel flow, are modified here by multiplying them by a normalised Fourier series:
	\begin{equation}
		\frac{1}{\sqrt{2\upi}},\quad  \frac{\sin(n t_1)}{\sqrt{\upi}},\quad  \frac{\cos(n t_1)}{\sqrt{\upi}}, \quad n = 1, 2, 3, \ldots.
	\end{equation}
	\item Constructing the inner product matrix $\mathsfbi{A}$, where its elements are expressed as:
	\begin{equation}
		A_{ij} = A(e_i,e_j) = \int_{0}^{1} \int_{0}^{2\upi} \int_{0}^{2\upi} e_i \cdot \mathcal{L}_{coo} e_j  \,\mathrm{d}t_1\, \mathrm{d}\theta\, \mathrm{d}z.
	\end{equation}
	\item Solving the weak formulation of the eigenvalue problem \cref{eq:eigenvalue problem} at a truncation degree:
	\begin{equation}
		\mathsfbi{A} \bm{\phi}_i = \lambda_i \bm{\phi}_i,
	\end{equation}
	where $\bm{\phi}_i$ is the vector of the coefficients of $f_i$.
\end{enumerate}

\vskip 10pt
Due to non-self-adjointness of $\mathcal{L}_{coo}$, the dual-eigenfunctions $\{f_i^{\star}\}_{i=1}^{\infty}$ are introduced for eigenfunction expansion, which satisfy biorthogonality relation with the eigenfunctions $\{f_i\}_{i=1}^{\infty}$ \citep{jiang_transient_2021}.
The biorthogonality relation is given by:
\begin{equation}
	\int_{0}^{1} \int_{0}^{2\upi} \int_{0}^{2\upi} f_i f_j^{\star} \,\mathrm{d}t_1\, \mathrm{d}\theta\, \mathrm{d}z = \delta_{ij}.
\end{equation}
This condition ensures mutual orthogonality between the eigenfunctions and their dual counterparts.
The dual eigenfunctions satisfy the following eigenvalue problem for the adjoint operator $\mathcal{L}_{coo}^{\star}$:
\begin{equation}
	\mathcal{L}_{coo}^{\star} f_i^{\star}  = \lambda_i f_i^{\star}. 
	\label{eq:dual eigenvalue problem}
\end{equation}

With the obtained eigenfunctions, dual-eigenfunctions, and eigenvalues, the local p.d.f. moments can be expanded into the following series \citep{barton_method_1983,jiang_transient_2021}:
\begin{equation}
	\hatP_n(z,\theta,t_1,t_0) = \sum_{i=1}^{\infty} p_{ni} \mathrm{e}^{-\lambda_i t_0} f_i(z,\theta,t_1),\quad n=0,1,\ldots, 
	\label{eq:moments expansion}
\end{equation}
where $p_{ni}$ are the expansion coefficients for local p.d.f. moments of each order.

For the first step in solving for the zeroth local p.d.f. moment $\hatP_0$, we find the expansion coefficients $p_{0i}$, using the initial condition \cref{eq:initial condition P0}:
\begin{align}
	p_{0i} = \int_{0}^{1} \int_{0}^{2\upi} \int_{0}^{2\upi} I_{ini}(z,\theta) f_i^{\star} \,\mathrm{d}t_1\, \mathrm{d}\theta\, \mathrm{d}z,
\end{align}
where $N$ denotes the truncation degree.
The higher-order local p.d.f. moments $\hatP_1$ and $\hatP_2$ can be successively solved with the series representations given in \citet[Appendix A]{jiang_transient_2021}.

With the solved transient moments in the two-time-variable system, we return to the moments in the single-time-variable system using
\begin{equation}
	P_n(z,\theta,t) = \hatP_n(z,\theta,t_0=t,t_1=\omega t)
\end{equation}
for straightforward definitions of the drift and dispersivity:
\begin{subequations}
\begin{equation}
	U_d \triangleq \frac{\mathrm{d} \left\langle P_1 \right\rangle_{z,\theta}}{\mathrm{d} t},
\end{equation}
\begin{equation}
	D_T \triangleq \frac{1}{2}\frac{\mathrm{d} \sigma^2}{\mathrm{d} t},
\end{equation}
\end{subequations}
where
\begin{equation}
	\sigma^2 = \left\langle P_2 \right\rangle_{z,\theta} - \left\langle P_1 \right\rangle_{z,\theta}^2
\end{equation}
is the mean square displacement of the cross-section-averaged concentration.

We validate our transient moments solutions by comparing with Brownian dynamics (BD) simulations, as described in Appendix \ref{app:BD}.
\Cref{fig:transient validation} presents a typical case of gyrotactic swimmers released as a uniform line source.
As shown, the results of method of moments and BD simulations are in good agreement.

In the method of moments, the components of the basis functions in each variable are truncated at the following degrees: $50$ for $z$, $12$ for $\theta$, and $4$ for $t_1$.
Additionally, $500$ pairs of eigenvalues and eigenfunctions are retained.
These truncation degrees are determined through independency tests, comparing the moments and cross-sectional concentration.

\begin{figure}
	\centerline{\includegraphics{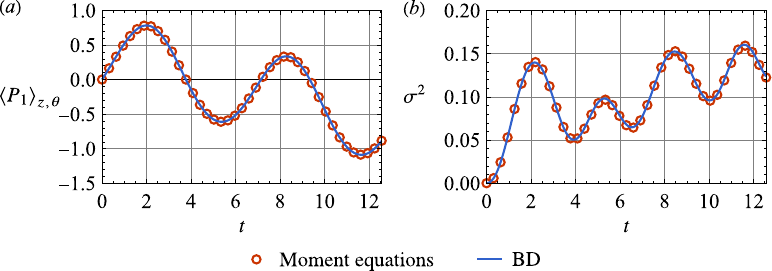}}
	\captionsetup{style=capcenter}
	\caption{
		Comparison of the results obtained from moments equations and BD simulations over the first two oscillation periods.
		($a$): First-order total moment $\left\langle P_1 \right\rangle_{z,\theta}$.
		($b$): Mean square displacement of the cross-section-averaged concentration $\sigma^2$.
		Note that quantities obtained with moments equations are expressed with the single original time variable $t$ using the substitutions $t_0 \to t$ and $t_1 \to \omega t$, and the hat symbol are simultaneously removed.
		Parameters: $\Pen_s=0.1$, $\Pen_f^s=0$, $\Pen_f^o=1$, $\alpha_0=0$, $\lambda=2.19$, $\omega=1$, $\Wo=1.72$, $I_{ini} =1/(2\upi)$.
	}
	\label{fig:transient validation}
\end{figure}

\section{GTD theory for the long-time asymptotic periodic dispersion}
\label{sec:GTD}
According to GTD theory for time-periodic dispersion problems \citep{brenner_macrotransport_1993}, the long-time asymptotic solution of zeroth-order local moment $P_0$, denoted by $P_0^{\infty}$, becomes periodic over the oscillation cycle.
Consequently, it is natural to seek the governing equation for the redefined zeroth-order local moment $\hatP_0^{\infty}$ within the two-time-variable formulation by neglecting the time-derivative with respect to $t_0$, while retaining the derivative with respect to $t_1$:
\begin{equation}
	\mathcal{L}_{coo} \hatP_0^{\infty} = 0. 
	\label{eq:governing equation asymptotic P0}
\end{equation}
\Cref{eq:governing equation asymptotic P0} can be solved using a Galerkin method with the same basis functions as described in \cref{sec:transient solutions}:
\begin{equation}
	\hatP_0^{\infty}(z,\theta,t_1) = \sum_{i=1}^{\infty} a_i e_i(z,\theta,t_1),
\end{equation}
which is equivalent to the expression in \cref{eq:moments expansion} for $n=0$ in the limit $t_0 \to \infty$.
We expect $\hatP_0^{\infty}$ to exhibit oscillatory behaviour for gyrotactic and elongated swimmers under the reflection boundary conditions, since their cross-sectional migration is predominantly influenced by the oscillatory shear.

\subsection{Long-time periodic drift}
The long-time asymptotic drift velocity can be directly evaluated as
\begin{equation}
	U_d^{\infty}(t_1) = \left\langle \hatP_0^{\infty} U_x \right\rangle_{z,\theta}, \label{eq:expression asymptotic drift}
\end{equation}
where the averaging is performed over the $(z,\theta)$ phase space only.
As a result, $U_d^{\infty}$ remains time-periodic.
To characterise the net transport over one oscillation cycle, we define the period-averaged asymptotic drift velocity as
\begin{equation}
	\overline{U_d^{\infty}} = \frac{1}{2\upi} \int_0^{2\upi} U_d^{\infty} \,\mathrm{d}t_1.
	\label{eq:expression Ud mean GTD}
\end{equation}

\subsection{Long-time period-averaged dispersivity}
We assume the solution for first-order total p.d.f. moment, further averaged over the oscillatory time variable $t_1$, to take the form
\begin{equation}
	\hatM_1(t_0) \sim \overline{U_d^{\infty}} t_0 + \overline{\left\langle b \right\rangle_{z,\theta}}+ \mathrm{e.d.t.}.\label{eq:assumed solution overline M1}
\end{equation}
\Cref{eq:assumed solution overline M1} can be interpreted as follows: from a period-averaged perspective, the centroid of the swimmer distribution moves with the period-averaged asymptotic drift velocity $\overline{U_d^{\infty}}$, offset by a constant term $\overline{\left\langle b \right\rangle_{z,\theta}}$, and accompanied by an exponential decay term in $t_0$, denoted as $\mathrm{e.d.t.}$.

Based on \cref{eq:assumed solution overline M1} and the relationship between $\hatP_1$ and $\hatM_1$ given in \cref{eq:Mn Pn}, we assume the solution for first-order local p.d.f. moment to take the form
\begin{equation}
	\hatP_1(z,\theta,t_1,t_0) \sim \hatP_0^{\infty} \overline{U_d^{\infty}} t_0 + b(z,\theta,t_1)  + \mathrm{e.d.t.}, \label{eq:assumed solution P1}
\end{equation}
where we define the scalar field
\begin{equation}
	B(z,\theta,t_1) = \frac{b}{\hatP_0^{\infty}} =  \lim_{t_0 \to \infty} \left( \frac{\hatP_1}{\hatP_0^{\infty}} - \hatM_1 \right) + \overline{\left\langle b \right\rangle_{z,\theta}}. \label{eq:B definition}
\end{equation}
Inspection of \cref{eq:B definition} reveals that $B(z,\theta,t_1)$ can be interpreted as the deviation between the mean streamwise position of swimmers located at $(z,\theta,t_1)$ and the overall mean streamwise position of all swimmers, with an addition of a constant $\overline{\left\langle b \right\rangle_{z,\theta}}$.
The scalar field $B(z,\theta,t_1)$, which is often called the `Brenner field', thus encapsulates all relevant information regarding the dispersion mechanisms \citep{brenner_macrotransport_1993,haugerud_solute_2022}.

Substituting \cref{eq:assumed solution P1} into the governing equation for $\hatP_1$ \cref{eq:governing P1}, we obtain the equation governing $b(z,\theta,t_1)$:
\begin{equation}
	\mathcal{L}_{coo} b = \hatP_0^{\infty} \left(U_x - \overline{U_d^{\infty}}\right).
\end{equation}
Rather than solving directly for $b$, we introduce a normalised Brenner field $b_N \triangleq b - \overline{\left\langle b \right\rangle_{z,\theta}} \hatP_0^{\infty}$, which satisfies
\begin{equation}
	\overline{\left\langle b_N \right\rangle_{z,\theta}} = 0.
	\label{eq:normalisation bN}
\end{equation}

The governing equation for $b_N$ takes the same form as that for $b$:
\begin{equation}
	\mathcal{L}_{coo} b_N = \hatP_0^{\infty} \left(U_x - \overline{U_d^{\infty}}\right).
	\label{eq:governing equation bN}
\end{equation}

The period-averaged dispersivity is defined as
\begin{equation}
	\overline{D_T^{\infty}} \triangleq \lim_{t_0 \to \infty} \frac{1}{2} \frac{\partial}{\partial t_0} \left(\hatM_2 - \hatM_1^2\right) .
\end{equation}
This expression can be simplified using \cref{eq:governing equations Mn}, yielding
\begin{equation}
	\overline{D_T^{\infty}} =  D_t + \lim_{t_0 \to \infty} \left(\overline{\left\langle U_x \hatP_1 \right\rangle_{z,\theta}} - \hatM_1 \frac{\partial \hatM_1}{\partial t_0} \right). \label{eq:DT simplified}
\end{equation}
Substituting \cref{eq:assumed solution overline M1} and \cref{eq:assumed solution P1} into \cref{eq:DT simplified}, we arrive at the final expression for the period-averaged dispersivity:
\begin{equation}
	\overline{D_T^{\infty}} = D_t 
	+ \overline{
		\left\langle
		U_x
		b_N
		\right\rangle_{z,\theta}
		}
		.
		\label{eq:expression DT mean GTD}
\end{equation}

The preceding derivation can be viewed as an extension of the framework presented by \citet[chapter 6]{brenner_macrotransport_1993}, which addresses scenarios without coupling between fluxes in the global and local spaces.
In contrast, our formulation incorporates the coupling between the global flux in the streamwise direction ($x$) and the local flux in the cross-section-orientaion space $(z,\theta)$, arising from the swimmers' self-propulsion.
In the absence of such coupling, the analysis simplifies considerably, as $\hatP_0^{\infty}$ becomes independent of $t_1$.

The long-time asymptotic period-averaged drift and dispersivity predicted by the GTD theory can be alternatively derived using a two-time-scale homogenisation method, as detailed in Appendix \ref{app:homogenisation}.
However, neither approach has, to the best of our knowledge, been extended to characterise the long-time asymptotic periodic (phase-dependent) dispersivity.

\subsection{Long-time periodic dispersivity}
To proceed further, we integrate \cref{eq:assumed solution P1} over the $(z,\theta)$ phase space to obtain
\begin{equation}
	\left\langle \hatP_1 \right\rangle_{z,\theta} \sim \left\langle \hatP_0^{\infty} \right\rangle_{z,\theta} \overline{U_d^{\infty}}  t_0 + \left\langle b \right\rangle_{z,\theta} + \mathrm{e.d.t.}.
	\label{eq:trial solution P1 integrated over z theta}
\end{equation}
Thus, the long-time asymptotic periodic dispersivity can be calculated using \cref{eq:trial solution P1 integrated over z theta,eq:governing equations Pn integrated over z theta}:
\begin{align}
	D_T^{\infty}(t_1) &\triangleq \frac{1}{2} \lim_{t_0 \to \infty}\frac{\mathrm{d}}{\mathrm{d} t}\left(\left\langle P_2 \right\rangle_{z,\theta} - \left\langle P_1 \right\rangle_{z,\theta}^2\right) \notag \\
	& = \frac{1}{2} \left( \frac{\partial \left\langle \hatP_2 \right\rangle_{z,\theta}}{\partial t_0} + \omega \frac{\partial  \left\langle \hatP_2 \right\rangle_{z,\theta}}{\partial t_1}\right)
	-\left \langle \hatP_1 \right\rangle_{z,\theta} \left\langle U_x \hatP_0^{\infty} \right\rangle_{z,\theta} \notag \\
	& = D_t + \left\langle \left(U_x - U_d^{\infty}\right) b_N \right\rangle_{z,\theta},\label{eq:asymptotic periodic DT}
\end{align}
where we have used the conservation condition in the ($z$,$\theta$) space:
\begin{equation}
	\left\langle \hatP_0^{\infty}\right\rangle_{z,\theta} = 1.
	\label{eq:hatP0 infty integrated over z theta}
\end{equation}
The above equation is deduced as follows:
First, we have
\begin{equation}
	\frac{\partial \hatP_0^{\infty}}{\partial t_0} = 0.
	\label{eq:asymptotic partial P0 partial t0}
\end{equation}
Integrating \cref{eq:asymptotic partial P0 partial t0} over $(z,\theta)$ phase space yields
\begin{equation}
	\frac{\partial \left\langle P_0^{\infty} \right\rangle_{z,\theta}}{\partial t_0} = 0.
	\label{eq:asymptotic partial P0 integrated over z theta partial t0}
\end{equation}
Substituting \cref{eq:asymptotic partial P0 integrated over z theta partial t0} into \cref{eq:governing equations Pn integrated over z theta} with $n=0$ results in
\begin{equation}
	\frac{\partial \left\langle P_0^{\infty} \right\rangle_{z,\theta}}{\partial t_1} = 0.
\end{equation}
Thus, $\left\langle P_0^{\infty} \right\rangle_{z,\theta}$ must be a constant.
Applying the normalisation condition \cref{eq:normalisation P}, we arrive at \cref{eq:hatP0 infty integrated over z theta}.

\begin{figure}
	\centerline{\includegraphics{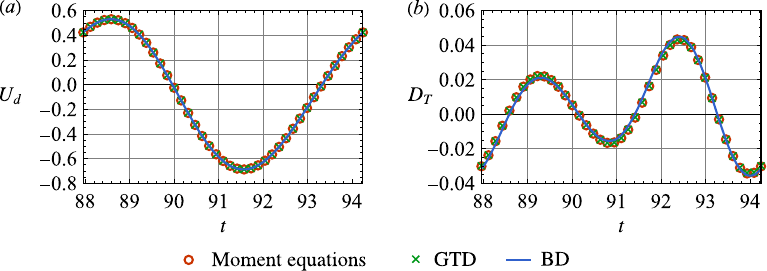}}
	\captionsetup{style=capcenter}
	\caption{
		Comparison of the results obtained from the moments equations, GTD, and BD simulations over an oscillation period long after the initial release ($t \in [14T,15T]$).
		($a$): Drift $U_d$.
		($b$): Dispersivity $D_T$.
		Note that quantities obtained with moment equations and GTD are expressed with the single original time variable $t$ using the substitutions $t_0 \to t$ and $t_1 \to \omega t$.
		The parameters are consistent with those used in \cref{fig:transient validation}.
	}
	\label{fig:asymptotic validation}
\end{figure}

We validate our GTD solutions for the periodic drift and dispersivity by comparing them with BD simulations and method of moments.
\Cref{fig:asymptotic validation} presents the same case as in \cref{fig:transient validation}, but with a time far from the initial release for the method of moments and BD simulations, as the GTD solutions are intended for long times.
Once again, we find good agreement between the results from GTD theory, the method of moments, and BD simulations.

\section{Results and discussion}
\label{sec:results}
We use a practical parameter space, as listed in \cref{tab:dimensional parameters},  referencing the model organisms \textit{Chlamydomonas augustae} with relaxation on the shape (Bretherton parameter $\alpha_0$) to encompass a broader range of micro-organisms, such as chain-forming micro-algae \citep{lovecchio_chain_2019} and bacteria \citep{ran_enhancing_2024}.
To clarify the how oscillatory shear modulates the dispersion of swimmers by influencing their local rotation, we also present the results for solute.
The solute diffusivity is set equal to the steady effective diffusivity of two-dimensional non-gyrotactic swimmers in an unbounded quiescent fluid, given by $\Pen_s^2/(n^2-n)+ D_t$ \citep{cates_when_2013}, where $n=2$ denotes the number of spatial dimensions.
For simplicity, we adopt a uniform line release:
\begin{equation}
	I_{ini} = \frac{1}{2\upi}.
	\label{eq:initial condition}
\end{equation} 

In the following, we present all results using the original single-time-variable system. 
Quantities derived using the method of moments, which formally depend on both $t_0$ and $t_1$, are now expressed with the single original time variable $t$.
Quantities obtained from the GTD theory, which periodically depend on $t_1$, are presented as functions of the original time variable modulo the oscillation period, i.e., $t \bmod T$.

\begin{table}
	\begin{center}
		\def~{\hphantom{0}}
		\begin{tabular}{lccc}
			Parameters  & Symbols  & Values/Ranges & Units\\
			Channel width & $W^{\ast}$ & $0.94$ & $\mathrm{cm}$\\
			Characteristic mean flow velocity of the steady pressure gradient & $U_s^{\ast}$ & $0$--$0.063$ &$\mathrm{cm} \, \mathrm{s}^{-1}$\\
			Characteristic mean flow velocity of the oscillatory pressure gradient & $U_o^{\ast}$ & $0$--$0.063$ &$\mathrm{cm} \, \mathrm{s}^{-1}$\\
			Kinematic viscosity of water & $\nu^{\ast}$ & $0.01$ & $\mathrm{cm}^2 \,  \mathrm{s}^{-1}$\\
			Angular frequency of oscillation & $\omega^{\ast}$ & $0.0067$--$0.67$ &$\mathrm{rad}\, \mathrm{s}^{-1}$ \\
			Swimming speed of swimmers & $V_s^{\ast}$ &$0.0063$ & $\mathrm{cm} \, \mathrm{s}^{-1}$\\
			Gravitactic reorientation time & $B^{\ast}$ & $3.4$--$\infty$ & $\mathrm{s}$\\
			Rotational diffusivity of swimmers & $D_r^{\ast}$ & $0.067$ & $\mathrm{rad}\,\mathrm{s}^{-1}$\\
			Translational diffusivity of swimmers & $D_t^{\ast}$ &$0$ & $\mathrm{cm}^2\,\mathrm{s}^{-1}$\\
			\mbox{}\\
			Flow P\'eclet number of the steady pressure gradient & $\Pen_f^s$ & $0$--$1$&\\
			Flow P\'eclet number of the oscillatory pressure gradient & $\Pen_f^o$ & $0$--$1$ &\\
			Dimensionless angular frequency of oscillation &$\omega$ &$0.1$--$10$ &\\
			Womersley number &$\Wo$ &$0.172$--$5.441$ &\\
			Swimming P\'eclet number & $\Pen_s$ & $0.1$ &\\
			Bretherton parameter & $\alpha_0$ &$0$--$1$ &\\
			Gravitactic bias parameter  & $\lambda$ & $0$--$2.19$ &\\
			Dimensionless translational diffusivity & $D_t$ &$0$ &
		\end{tabular}
		\caption{
			Parameters used for swimmers in this work.
			The values of $V_s^{\ast}$, $B^{\ast}$, and $D_r^{\ast}$ are based on the model organisms \textit{Chlamydomonas augustae} (data primarily sourced from \citet{pedley_new_1990} and \citet{hwang_bioconvection_2014,hwang_stability_2014}).
			Note that, since constant values of $W^{\ast}$, $\nu^{\ast}$, and $D_r^{\ast}$ are used, the Womersley number, $\Wo$, is uniquely determined by the relation $\Wo = 1.72 \sqrt{\omega}$.
		}
		\label{tab:dimensional parameters}
	\end{center}
\end{table}

\subsection{Effects of swimming ability and oscillatory flow strengths}
\begin{figure}
	\centerline{\includegraphics{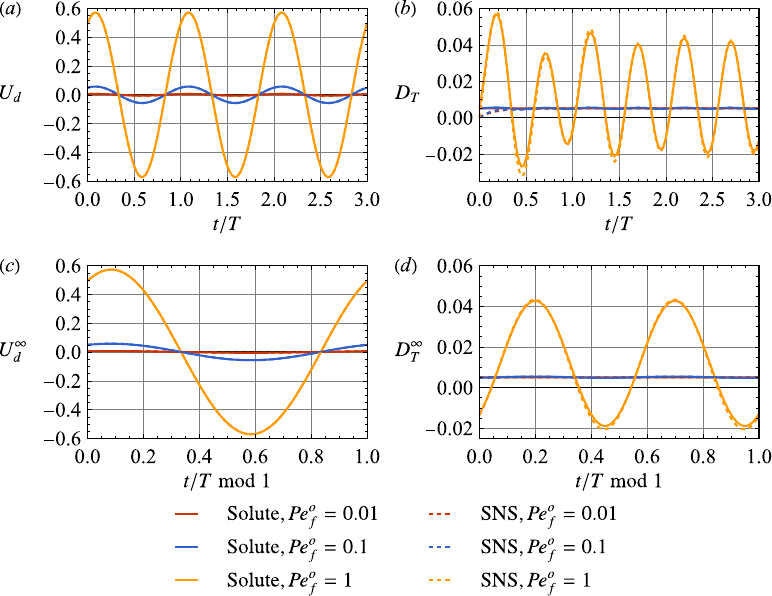}}
	\captionsetup{style=capcenter}
	\caption{
		($a$,$b$): Transient drift $U_d$ and dispersivity $D_T$ of solute and spherical non-gyrotactic swimmers (SNS) over the first three periods following a uniform line release for several oscillatory flow P\'eclet numbers $\Pen_f^o$.
		($c$,$d$): Long-time asymptotic periodic drift $U_d^{\infty}$ and dispersivity $D_T^{\infty}$ of solute and SNS over one period for several oscillatory flow P\'eclet numbers $\Pen_f^o$.
		Parameters for flow: $\Pen_f^s=0$, $\omega=1$, $\Wo=1.72$.
		Parameters for solute: $\Pen_s=0$, $\alpha_0=0$, $\lambda=0$, $D_t=0.005$.
		Parameters for SNS: $\Pen_s=0.1$, $\alpha_0=0$, $\lambda=0$, $D_t=0$.
	}
	\label{fig:temporal evolution - effects of swimming and oscillatory flow strength}
\end{figure}

In this subsection, we investigate the effects of swimming ability, $\Pen_s$, and oscillatory flow strengths, $\Pen_f^o$, on both the transient and long-time asymptotic dispersion.
We focus on spherical non-gyrotactic swimmers (SNS), where the distribution in $(z,\theta)$ space remains uniform for the uniform line release \cref{eq:initial condition} under reflective boundary conditions, i.e., $\hatP_0(z,\theta,t_1) = 1/(2\upi)$ \citep{jiang_transient_2021}.
As a result, both the transient drift, $U_d$, and the long-time asymptotic periodic drift, $U_d^{\infty}$, are identical for solute and SNS, perfectly tracking the instantaneous mean flow speed, as seen in figures \ref{fig:temporal evolution - effects of swimming and oscillatory flow strength}($a$) and \ref{fig:temporal evolution - effects of swimming and oscillatory flow strength}($c$).

As shear disrupts the cross-sectional migration of swimmers in an oscillatory manner, their dispersivity behaves differently from solute, which undergo simple molecular diffusion across the cross-section.
Figures \ref{fig:temporal evolution - effects of swimming and oscillatory flow strength}($b$) and \ref{fig:temporal evolution - effects of swimming and oscillatory flow strength}($d$) show the subtle differences in the time evolution of dispersivity of SNS compared to solute.
Initially, following release, $D_T$ of solute starts at its molecular diffusivity, $D_t=0.005$, whereas $D_T$ of SNS begins at zero and gradually approaches that of solute.
This can be attributed to the fact that the mean square displacement of SNS exhibit Brownian diffusive behaviours with diffusivity $D_t$ (set to zero in our work) at very short time scales in the absence of flow \citep{bechinger_active_2016}.
Beyond this initial stage, the difference in $D_T$ between solute and SNS becomes most pronounced at $\Pen_f^o=1$, where the flow dominates over swimming ($\Pen_f^o \gg \Pen_s$); in other cases, the oscillatory behaviour of $D_T$ remains relatively weak.
The most notable differences appear at the peaks and troughs of the oscillatory $D_T$: $D_T$ of SNS exceeds that of solute at the peaks but falls below it at the troughs.
This phenomenon can be interpreted as follows: when the flow develops strongly in one direction, SNS are dispersed more effectively due to suppressed cross-streamline migration under Jeffery orbits, which is reminiscent of the inverse dependence of Taylor dispersivity on molecular diffusivity in steady flows.
However, when the flow reverses direction, SNS tend to return more closely to their initial position compared to solute, as their ability to diffuse across the cross-section is limited.
For both solute and SNS, the emergence of negative $D_T$ arises from our use of the widely adopted definition of $\sigma^2$, which involves averaging across the cross-section before computing the mean square displacement, rather than the alternative definition proposed by \citet{yasuda_longitudinal_1984}, which calculates mean square displacement at each streamline first.

At long times, $D_T^{\infty}$ becomes periodic for both solute and SNS.
Furthermore, \cref{fig:temporal evolution - effects of swimming and oscillatory flow strength}($d$) reveals that the actual period of $D_T^{\infty}$ is $T/2$, due to the two symmetrical dispersion cycles that occur when the mean flow velocity points in opposite directions, which is in line with the preliminary conjecture of \citet{chatwin_longitudinal_1975} and the more formal results given by \citet{yasuda_longitudinal_1984}.
The two dispersion cycles are identical in the variations of $D_T^{\infty}$, which is independent of phase of the initial release.
Each cycle can be further separated into two intervals: one with positive $D_T^{\infty}$, corresponding to the attenuation of mean flow velocity in one direction, and one with negative $D_T^{\infty}$, corresponding to the strengthening of mean flow velocity in the opposite direction.
Moreover, the integration of $D_T^{\infty}$ over time during the positive interval outweighs the integration during the negative interval, leading to a positive period-averaged dispersivity.

\begin{figure}
	\centerline{\includegraphics{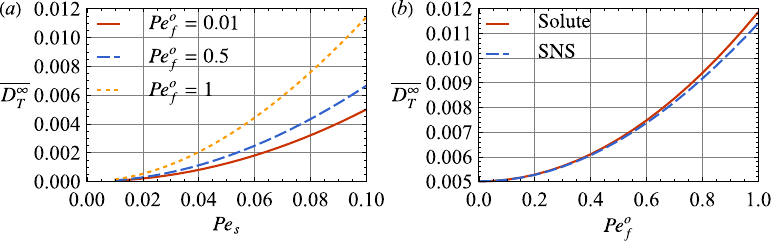}}
	\captionsetup{style=capcenter}
	\caption{
		Long-time asymptotic period-averaged dispersivity $\overline{D_T^{\infty}}$ as functions of ($a$) the swimming P\'eclet number $\Pen_s$ and ($b$) the oscillatory flow P\'eclet number $\Pen_f^o$.
		Parameters for flow: $\Pen_f^s=0$, $\omega=1$, $\Wo=1.72$.
		Parameters in ($a$): $\alpha_0=0$, $\lambda=0$, $D_t=0$.
		Parameters in ($b$): $\Pen_s=0$, $\alpha_0=0$, $\lambda=0$, $D_t=0.005$ (solute); $\Pen_s=0.1$, $\alpha_0=0$, $\lambda=0$, $D_t=0$ (SNS).
	}
	\label{fig:mean value - effects of swimming and oscillatory flow strength}
\end{figure}

The long-time asymptotic period-averaged dispersivity, $\overline{D_T^{\infty}}$, is plotted in \cref{fig:mean value - effects of swimming and oscillatory flow strength} as functions of $\Pen_s$ and $\Pen_f^o$.
Note that the period-averaged drift, $\overline{U_d^{\infty}}$, is always zero for both solute and SNS due to the uniform distribution in the $(z,\theta)$ space, so it is not shown here fore conciseness.
Increasing trends of $\overline{D_T^{\infty}}$ with $\Pen_s$ and $\Pen_f^o$ are observed, indicating the positive influences of swimming ability and oscillatory flow strength on the dispersion at the period-averaged level over asymptotically long times.
Furthermore, in \cref{fig:mean value - effects of swimming and oscillatory flow strength}($b$), we observe that as $\Pen_f^o$ increases, $\overline{D_T^{\infty}}$ of solute exceed that of SNS, indicating a net weakening effect of motility on dispersion in oscillatory flow..
However, we speculate that this trend may not hold across a broader range of oscillation frequency $\omega$.

\subsection{Effects of oscillation frequency}
\begin{figure}
	\centerline{\includegraphics{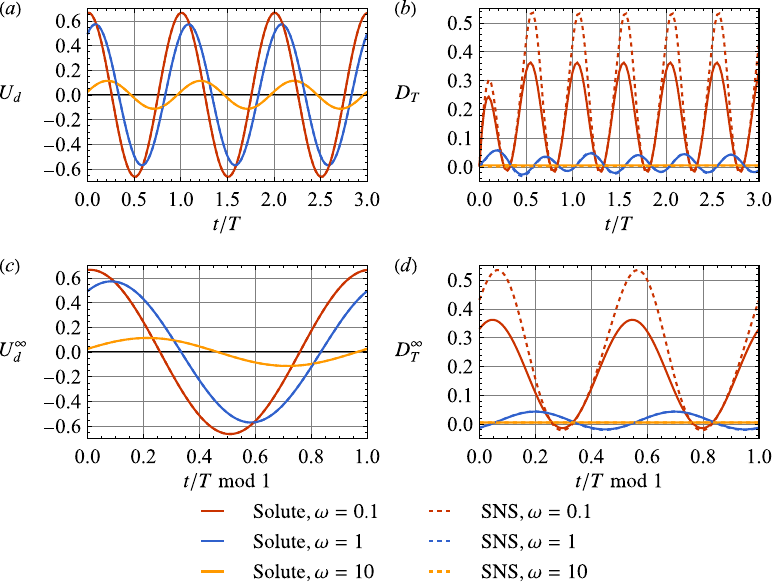}}
	\captionsetup{style=capcenter}
	\caption{		
		($a$,$b$): Transient drift $U_d$ and dispersivity $D_T$ of solute and spherical non-gyrotactic swimmers (SNS) over the first three periods following a uniform line release for several oscillation frequency $\omega$.
		($c$,$d$): Long-time asymptotic periodic drift $U_d^{\infty}$ and dispersivity $D_T^{\infty}$ of solute and SNS over one period for several oscillation frequency $\omega$.
		Parameters for solute: $\Pen_s=0$, $\alpha_0=0$, $\lambda=0$, $D_t=0.005$.
		Parameters for SNS: $\Pen_s=0.1$, $\alpha_0=0$, $\lambda=0$, $D_t=0$.
	}
	\label{fig:temporal evolution - effects of oscillation frequency}
\end{figure}

In this subsection, we examine the previously conjecture regarding the influence of the oscillation frequency $\omega$ on active dispersion in oscillatory flows.
\Cref{fig:temporal evolution - effects of oscillation frequency} presents the transient and long-time asymptotic periodic drift and dispersivity for several values of $\omega$, with comparisons between solute and SNS.
While both $U_d$ and $U_d^{\infty}$ continue to follow the instantaneous mean flow velocity, their magnitudes decrease and their phases shift rightward as $\omega$ increases.
This behaviour reflects the increasing influence of fluid inertia, characterised by the Womersley number $\Wo$.

\begin{figure}
	\centerline{\includegraphics{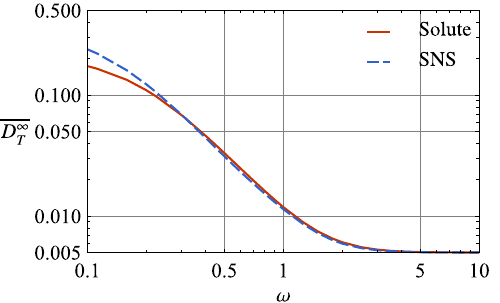}}
	\captionsetup{style=capcenter}
	\caption{
		Long-time asymptotic period-averaged dispersivity $\overline{D_T^{\infty}}$ as a function of oscillation frequency $\omega$.
		The parameters for solute and SNS are consistent with those used in \cref{fig:temporal evolution - effects of oscillation frequency}.
	}
	\label{fig:mean value - effects of oscillation frequency}
\end{figure}

The amplitude of dispersivity oscillations diminishes more noticeably with increasing $\omega$, due to the quadratic dependence of dispersivity on the flow velocity.
At the lowest oscillation frequency considered ($\omega=0.1$), the transient dispersivity of SNS is significantly larger than that of solute, except for at the troughs, where the two values nearly converge.
At $\omega=1$, the dispersivity curves of SNS and solute nearly overlap, with only minor differences at the peaks and troughs.
At the highest oscillation frequency ($\omega=10$), the dispersivity of SNS is consistently smaller than that of solute.
The complex influence of oscillation frequency is illustrated more clearly from a long-time asymptotic period-averaged perspective.
As shown in \cref{fig:mean value - effects of oscillation frequency}, although the dispersivity $\overline{D_T^{\infty}}$ for both SNS and solute declines rapidly with increasing $\omega$, $\overline{D_T^{\infty}}$ for SNS exceeds that of solute at relatively low oscillation frequencies $\omega < 0.3$, but becomes smaller at higher oscillation frequencies $\omega > 0.3$.
Since $\omega$ is defined as the ratio of dimensional oscillation frequency to the rotational diffusivity, this observation underscores the intricate coupling between swimming dynamics and oscillatory shear in determining the dispersion of SNS.

\subsection{Effects of superimposed steady component on dispersion in oscillatory flows}
\begin{figure}
	\centerline{\includegraphics{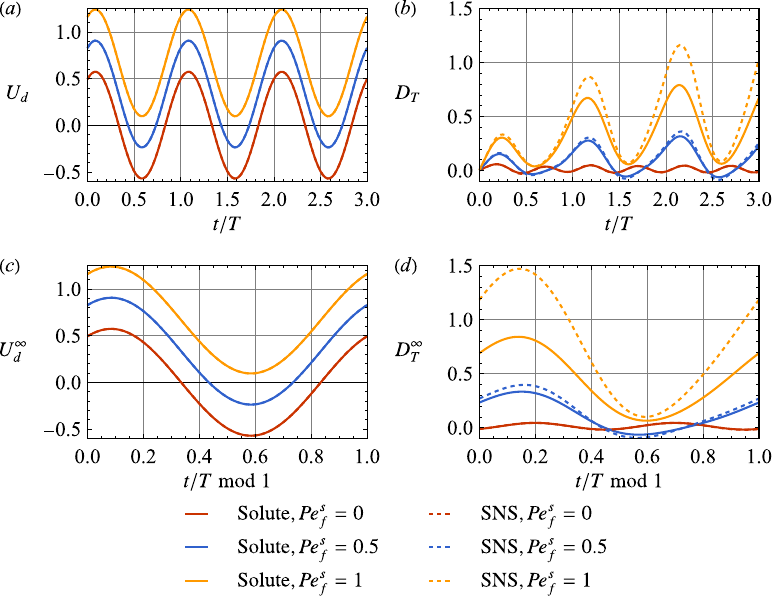}}
	\captionsetup{style=capcenter}
	\caption{		
		($a$,$b$): Transient drift $U_d$ and dispersivity $D_T$ of solute and spherical non-gyrotactic swimmers (SNS) over the first three periods following a uniform line release for several steady flow P\'eclet numbers $\Pen_f^s$.
		($c$,$d$): Long-time asymptotic periodic drift $U_d^{\infty}$ and dispersivity $D_T^{\infty}$ of solute and SNS over one period for several steady flow P\'eclet numbers $\Pen_f^s$.
		Parameters for flow: $\Pen_f^o=1$, $\omega=1$, $\Wo=1.72$.
		Parameters for solute: $\Pen_s=0$, $\alpha_0=0$, $\lambda=0$, $D_t=0.005$.
		Parameters for SNS: $\Pen_s=0.1$, $\alpha_0=0$, $\lambda=0$, $D_t=0$.
	}
	\label{fig:temporal evolution - effects of steady flow}
\end{figure}

This subsection investigates the effects of a superimposed steady component on dispersion in oscillatory flows.
\Cref{fig:temporal evolution - effects of steady flow} presents the transient and long-time asymptotic variations of drift and dispersivity for several steady flow P\'eclet number $\Pen_f^s$, with a fixed oscillatory flow P\'eclet number $\Pen_f^o=1$.
The combination of oscillatory flow and steady flow does not alter the uniform distribution in the $(z,\theta)$ space; therefore the drift remains equivalent to the instantaneous mean flow velocity, contributed by both the steady and oscillatory components.
In terms of dispersivity, the presence of a steady flow leads to an increased oscillation amplitude --- specifically, the peaks rise more than the troughs fall.
Additionally, only a single dispersion cycle occurs over an asymptotic oscillation period, unlike the two cycles observed previously when $\Pen_f^s = 0$.

\begin{figure}
	\centerline{\includegraphics{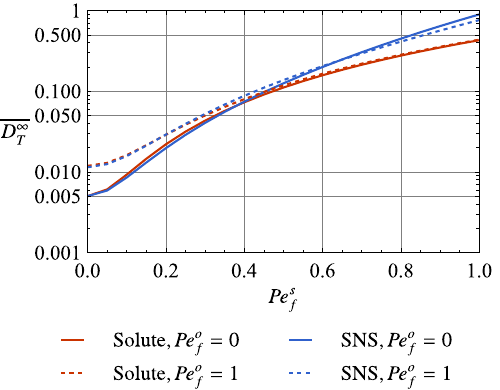}}
	\captionsetup{style=capcenter}
	\caption{
		Long-time asymptotic period-averaged dispersivity $\overline{D_T^{\infty}}$ plotted against the steady flow P\'eclet number $\Pen_f^s$.
		Parameters for flow: $\omega=1$, $\Wo=1.72$.
		The parameters for solute and SNS are consistent with those used in \cref{fig:temporal evolution - effects of steady flow}.
	}
	\label{fig:mean value - effects of steady flow}
\end{figure}

The introduction of a steady component also results in a more complex variation in the dispersivity of SNS.
As illustrated in \cref{fig:mean value - effects of steady flow}, there is a dual effect of an oscillatory component on the dispersion of SNS when a steady component is present: at small $\Pen_f^s$, oscillation enhances dispersion, whereas at large $\Pen_f^s$, oscillation inhibits it.
In contrast, for solute, the addition of oscillation consistently enhances dispersion compare with purely steady flow.

\subsection{Effects of gyrotaxis and elongation}
\begin{figure}
	\centerline{\includegraphics{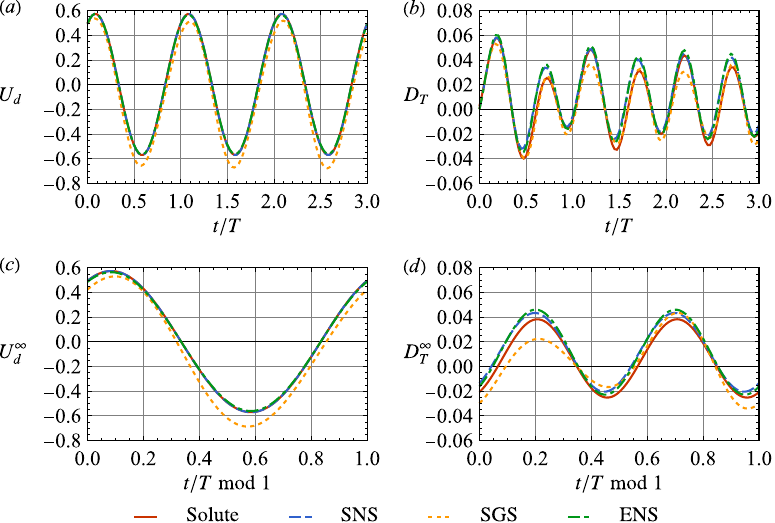}}
	\captionsetup{style=capcenter}
	\caption{		
		($a$,$b$): Transient drift $U_d$ and dispersivity $D_T$ of solute, spherical non-gyrotactic swimmers (SNS), spherical gyrotactic swimmers (SGS), and elongated non-gyrotactic swimmers (ENS) over the first three periods following a uniform line release.
		($c$,$d$): Long-time asymptotic periodic drift $U_d^{\infty}$ and dispersivity $D_T^{\infty}$ of solute, SNS, SGS, and ENS over one period.
		Parameters for solute: $\Pen_s=0$, $\alpha_0=0$, $\lambda=0$, $D_t=0.005$.
		Parameters for SNS: $\Pen_s=0.1$, $\alpha_0=0$, $\lambda=0$, $D_t=0$.
		Parameters for SGS: $\Pen_s=0.1$, $\alpha_0=0$, $\lambda=2.19$, $D_t=0$.
		Parameters for ENS: $\Pen_s=0.1$, $\alpha_0=1$, $\lambda=0$, $D_t=0$.
	}
	\label{fig:temporal evolution - effects of gyrotaxis and elongation}
\end{figure}

\begin{figure}
	\centerline{\includegraphics{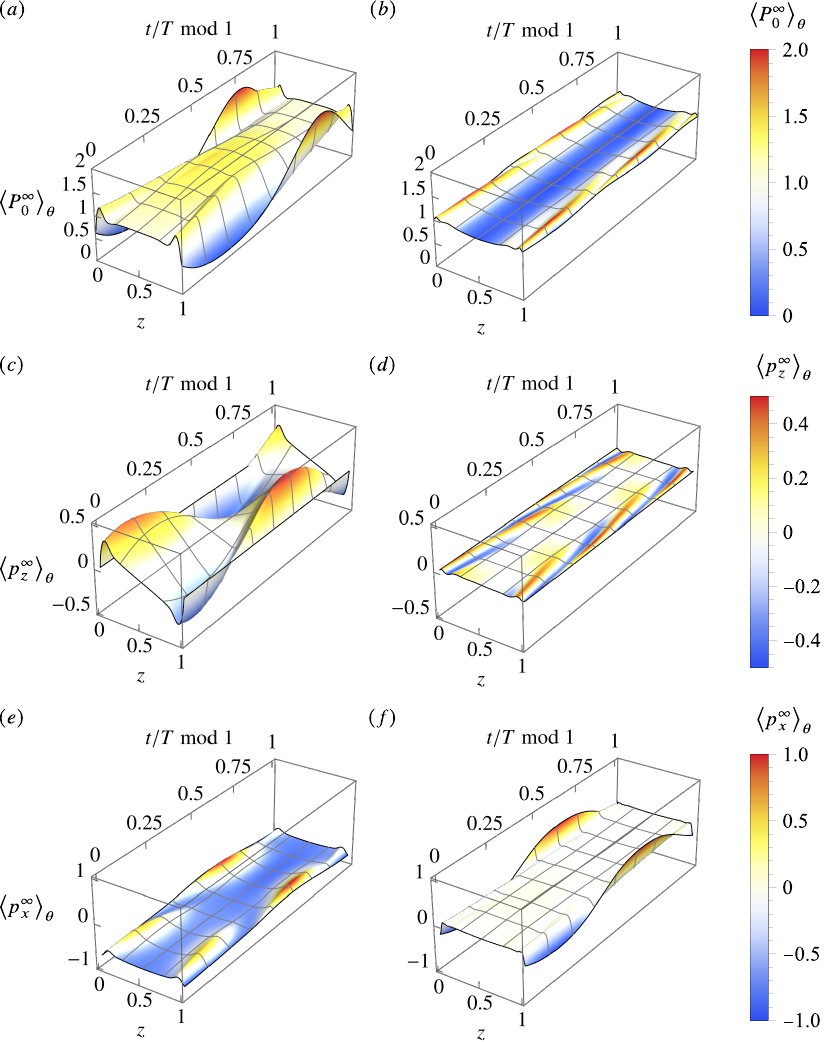}}
	\captionsetup{style=capcenter}
	\caption{		
		Long-time asymptotic periodic cross-sectional distribution $\left\langle P_0^{\infty} \right\rangle_{\theta}\triangleq \int_0^{2\upi} P_0^{\infty} \, \mathrm{d} \theta$, local mean swimming direction component along the cross-section $\left\langle p_z^{\infty} \right\rangle_{\theta} \triangleq \left(\int_0^{2\upi} P_0^{\infty} \sin\theta \, \mathrm{d} \theta\right)\left/ \left(\int_0^{2\upi} P_0^{\infty} \, \mathrm{d} \theta\right)\right.$ and along the streamwise direction $\left\langle p_x^{\infty} \right\rangle_{\theta} \triangleq \left(\int_0^{2\upi} P_0^{\infty} \cos\theta \, \mathrm{d} \theta\right)\left/\left(\int_0^{2\upi} P_0^{\infty} \, \mathrm{d} \theta\right)\right.$ over one period.
		($a$,$c$,$e$): Spherical gyrotactic swimmers (SGS).
		($b$,$d$,$f$): Elongated non-gyrotactic swimmers (ENS).
		Flow parameters: $\Pen_f^s=0$, $\Pen_f^o=1$, $\omega=1$, $\Wo=1.72$.
		Particles parameters are consistent with those used in \cref{fig:temporal evolution - effects of gyrotaxis and elongation}.
	}
	\label{fig:3D con - effects of gyrotaxis and elongation}
\end{figure}

This subsection discusses the effects of gyrotaxis and elongation on oscillatory dispersion.
\Cref{fig:temporal evolution - effects of gyrotaxis and elongation} compares the transient and long-time asymptotic periodic drift and dispersivity for solute, spherical non-gyrotactic swimmers (SNS), spherical gyrotactic swimmers (SGS), and elongated non-gyrotactic swimmers (ENS).
While ENS slightly always drift slower than the instantaneous mean flow, SGS exhibit more pronounced variations --- drifting faster during upward flow and slower during downward flow.
This behaviour is not attributed to the well-known response of SGS to shear flow: in steady conditions, SGS typically undergo gyrotactic focusing near the centre or walls of the channel for downwelling or upwelling flows, resulting in concentration accumulations significantly exceeding the mean and causing enhanced drift during downward flow and reduced drift during upward flow.
However, in oscillatory flows, the cross-sectional concentration distribution of SGS cannot immediately follows the time-varying shear profile due to the constraints of weak swimming strength ($\Pen_s=0.1$) and limited gyrotactic reorientation.
As shown in \cref{fig:3D con - effects of gyrotaxis and elongation}($a$), the cross-sectional concentration profile of SGS remains relatively flat in the central region over a period, with notable gradients only occurring near the walls --- starkly contrasting with the steady-flow case.
\Cref{fig:3D con - effects of gyrotaxis and elongation}($c$) further presents the long-time asymptotic periodic local mean swimming direction component along the cross-section $\left\langle p_z^{\infty} \right\rangle_{\theta}$ of SGS over a period.
Significant gradients in $\left\langle p_z^{\infty} \right\rangle_{\theta}$ are again confined to the near-wall region, suggesting corresponding localised concentration variations.

A full sampling of the non-uniform shear field is crucial to the shear-induced trapping of ENS, which typically requires a long time and streamwise distances \citep{rusconi_bacterial_2014,vennamneni_shearinduced_2020}.
Moreover, the concentration enhancement in steady flow remains of the same order as the mean concentration.
As shown in \cref{fig:3D con - effects of gyrotaxis and elongation}($b$), the cross-sectional concentration profiles of ENS are even more uniform, with only weak gradients near the walls.
Figures \ref{fig:3D con - effects of gyrotaxis and elongation}($d$) and \ref{fig:3D con - effects of gyrotaxis and elongation}($f$) present the long-time asymptotic periodic local mean cross-sectional and streamwise swimming direction components over a period, $\left\langle p_z^{\infty} \right\rangle_{\theta}$ and $\left\langle p_x^{\infty} \right\rangle_{\theta}$, respectively, further confirming the weak accumulation and alignment of ENS in oscillatory flows.

The stronger response of SGS in terms of concentration distribution and mean swimming direction to oscillatory flow leads to its distinct dispersivity compared with other types of particle.
In contrast, the relatively uniform and weak responses of ENS result in dispersivity characteristics more similar to those of solute and SNS, as shown in figures \ref{fig:temporal evolution - effects of gyrotaxis and elongation}($b$) and \ref{fig:temporal evolution - effects of gyrotaxis and elongation}($d$).

\begin{figure}
	\centerline{\includegraphics{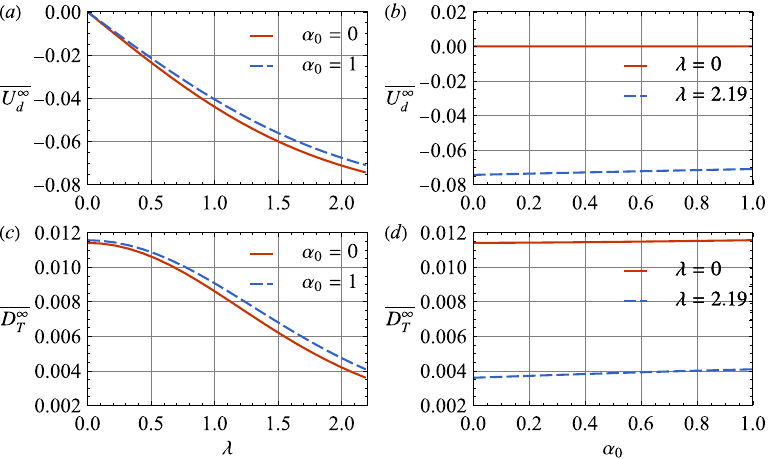}}
	\captionsetup{style=capcenter}
	\caption{
		Long-time asymptotic period-averaged drift $\overline{U_d^{\infty}}$ and dispersivity $\overline{D_T^{\infty}}$ as functions of ($a$,$c$) gravitactic bias parameter and ($b$,$d$) Bretherton parameter $\alpha_0$.
		Flow parameters are consistent with those used in \cref{fig:3D con - effects of gyrotaxis and elongation}.
	}
	\label{fig:mean value - effects of gyrotaxis and elongation}
\end{figure}

\Cref{fig:mean value - effects of gyrotaxis and elongation} presents the long-time asymptotic period-averaged drift $\overline{U_d^{\infty}}$ and dispersivity $\overline{D_T^{\infty}}$ as functions of the gravitactic bias parameter $\lambda$ and Bretherton parameter $\alpha_0$.
Gyrotaxis is found to induce a negative mean drift velocity, which can be explained by the steady upwards streamwise alignment observed in \cref{fig:3D con - effects of gyrotaxis and elongation}($e$).
It also weakens the overall dispersivity.
Elongation, on the other hand, does not produce a net drift for non-gyrotactic swimmers but slightly enhances $\overline{U_d^{\infty}}$ for gyrotactic swimmers.
For both swimmer types, elongation leads to a modest increase in $\overline{D_T^{\infty}}$.

\section{Concluding remarks}
\label{sec:concluding remarks}
This work combines a two-time-variable expansion for the transient dispersion and the GTD theory for the long-time asymptotic periodic dispersion to investigate the Taylor--Aris dispersion of active particles in oscillatory channel flows.
The two-time-variable expansion which is introduced to capture the periodicity in transient evolution to the Taylor regime due to oscillation of the flow, gains deeper interpretation through the lens of GTD theory: in the long-time asymptotic limit, the dispersion problem simplifies to a periodic problem governed solely by the oscillatory time variable $t_1=\omega t$.

Traditional approximating models based on orientation--position separation, such as two-step GTD model \citep{bearon_spatial_2011}, typically assume a quasi-steady and quasi-uniform shear.
These models solve the equilibrium orientation pointwise and then compute the drift velocity and diffusivity tensor in position space.
However, such assumptions break down in flows with strong spatial inhomogeneity \citep{bearon_spatial_2011,jiang_dispersion_2020,wang_gyrotactic_2022}.
Recently, \citet{caldag_fine-tuning_2025} also demonstrated the failure of the two-step GTD model at high oscillation frequencies, where flow conditions deviate significantly from quasi-steady assumption.
At the other end of the oscillation spectrum, while these models may provide accurate predictions for very low-frequency oscillations, they often require prohibitively long simulations and large computational domains to reach the asymptotic Taylor regime \citep{caldag_fine-tuning_2025}. 
In contrast, our transient dispersion framework efficiently captures key statistical features in terms of  moments, without resolving the full concentration field explicitly.
This makes it particularly suitable for low-frequency oscillatory flows as well.
Therefore, compared to traditional approximating models, our method accommodates a broader range of oscillation frequencies $\omega$, offering a promising tool for understanding and controlling dispersion of active particles in oscillatory flows.

Employing both approaches, we conduct detailed analyses of swimmer dispersion following release from a uniform line source, considering potential additional effects from gyrotaxis and elongation.
For spherical non-gyrotactic swimmers, although their zeroth-order moment remains uniform --- similar to that of solute --- the presence of oscillatory flow can either enhance or reduce their dispersion relative to solute with comparable molecular diffusivity.
The enhancement, which occurs at low frequencies, is attributed to the swimmers' reduced effective ability to migrate across streamline, as described by the Jeffery equation, which captures for the linear dependence of rotation on shear rate for spherical particles.
In contrast, the reduction at high frequencies is due to the nearly resetting of swimmers' streamwise positions after each short oscillation cycle, a consequence of limited cross-sectional homogenising.
The case of superimposing a steady component onto an oscillatory flow is also investigated, revealing a dual effect of oscillation on swimmer dispersion, in contrast to the monotonic enhancement observed for solute.
Two potential terms in the Jeffery orbit with bias to gravity and rate-of-strain --- gyrotaxis and elongation --- are subsequently considered.
While both induce non-uniform cross-sectional distributions, only gyrotaxis significantly alters the dispersion characteristics, owing to its higher sensitivity to shear.
By contrast, although elongation allows swimmers to respond to shear gradients, their trapping in high-shear regions requires extended time and distance.
Consequently, in an oscillatory environment, where such sustained exposure is absent, shear-induced trapping fails to develop appreciably as in steady flows and thus has a limited effect on dispersion.

This work assumes an extremely dilute suspension, neglecting swimmer--swimmer interaction.
In practice, for naturally buoyant swimmers (typically 5\% denser than water), buoyancy-driven effects dominate swimmer--swimmer interactions in steady gyrotactic plumes \citep{fung_bifurcation_2020}.
In oscillatory flows, the development of gyrotactic plumes is hindered, weakening both buoyancy effects and far-field interactions.
If buoyancy--flow coupling becomes non-negligible, local swimmer accumulation may alter density distributions, resulting in a more pronounced phase lag between flow velocity and pressure gradient.
Additionally, considering far-field interactions for puller-type swimmers may increase the effective viscosity \citep{saintillan_rheology_2018}, further amplifying this phase lag.

Swimmer--boundary interaction represent another important factor that can influence active particle dispersion in oscillatory flows.
The reflective boundary conditions employed in the current study do not capture wall accumulation effects, which may lead to an underestimation of near-wall concentrations.
Previous studies have shown that wall-accumulating swimmers exhibit reduced dispersivity in moderate steady flows \citep{jiang_dispersion_2019}.
In contrast, in strong steady flows, the dispersivity approaches that of non-accumulating swimmers due to the suppression of wall polarisation by shear.
In oscillatory flows, we speculate that wall accumulation could be more pronounced than in steady flows, owing to the reduced shear rate near the wall.
Such enhanced accumulation may further suppress dispersion.

The current work can be extended to other periodic processes beyond oscillatory flows.
Recent studies \citep{omori_rheotaxis_2022,walker_emergent_2022} suggest that the periodic variations in swimming speed and body shape may explain the experimentally observed rheotaxis and centreline migration of swimmers.
Additionally, based on this work, a multi-time-variable expansion can be employed to address the situations when multiple time-periodic processes coexist, each with distinct base frequencies.
Further development could involve constructing a period-resolved transport model capturing the full regimes of concentration evolution.
Although applying the long-time asymptotic period-averaged one-dimensional dispersion equation \cref{eq:effective 1D dispersion equation} avoids the singularity associated with transient negative diffusivity, it resolves none of the cross-sectional concentration, the characteristics within the period, and the transient dynamics.
A potential improvement would be to approximate the concentration distribution by fully utilising the transient concentration moments at each streamline, such as applying the Edgeworth expansion \citep{chatwin_approach_1970,wang_contaminant_2017,guo_transient_2020,li_environmental_2023,guan_streamwise_2024} or Gill's generalised dispersion model \citep{gill_analysis_1967,gill_note_1967}, both of which have been shown to be efficient in solute dispersion.

\begin{bmhead}[Acknowledgements.]
We are grateful to Prof. G.Q. Chen for his valuable discussions during the initial conceptualisation of this work.
\end{bmhead}

\begin{bmhead}[Funding.]
This work is supported by the National Key R\&D Program of China (Grant No. 2024YFC3210902), the IWHR Research and Development Support Program (Grant No. HY0199A112021), and the Independent Research Project of State Key Laboratory of Water Cycle and Water Security in River Basin (Grant No. SKL2024TS12).
B.W. acknowledges support from the Youth Talent Lifting Project of the Department of Hydraulics, IWHR (Grant No. HY121003A0012025).
\end{bmhead}

\begin{bmhead}[Declaration of Interests.]
The authors report no conflict of interest.
\end{bmhead}

\appendix
\begin{appen}
	
\section{BD simulations}
\label{app:BD}
We use BD simulations to validate our solutions for the transient moment equations and the long-time asymptotic GTD theory.
The simulations are performed using the Langevin equation for the single-particle motion:
\begin{subequations}
	\begin{equation}
		\mathrm{d} x = U \mathrm{d} t + \Pen_s \cos \theta \mathrm{d} t + \sigma_x \mathrm{d} W_x, 
		\label{eq:dx}
	\end{equation}
	\begin{equation}
		\mathrm{d} z = \Pen_s \sin\theta \mathrm{d} t + \sigma_z \mathrm{d} W_z,
		\label{eq:dz}
	\end{equation}
	\begin{equation}
		\mathrm{d} \theta =  \dot{\theta} \mathrm{d} t + \sigma_r \mathrm{d} W_{\theta}.
		\label{eq:dtheta}
	\end{equation}
\end{subequations}
Here $W_x$, $W_z$, and $W_{\theta}$ are independent standard Brownian motions, $\sigma_x = \sigma_z = \sqrt{2 D_t}$, and $\sigma_r = \sqrt{2}$.

After discretising \cref{eq:dx,eq:dz,eq:dtheta} using the Euler--Maruyama forward scheme, we simulate $10^5$ trajectories of swimmers with a time step of $\Delta t = 10^{-4} T$.
For swimmers outside the boundary $z \in [0,1]$, a reflection operation on both position and orientation is performed with respect to the adjacent wall, ensuring consistency with the reflective boundary conditions given in \cref{eq:reflective boundary conditions P,eq:reflective boundary conditions pPpz}.
Concentration moments and distributions are then extracted to validate the theoretical models.

\section{Two-time-scale homogenisation for the long-time asymptotic period-averaged drift and dispersivity}
\label{app:homogenisation}
In this appendix we present a two-time-scale homogenisation method for determining the long-time asymptotic period-averaged drift and dispersivity.
The single-time-scale homogenisation method \citep[chapter 12]{pavliotis_multiscale_2008} has been widely applied to dispersion in steady flows \citep{pavliotis_multiscale_2005,chen_shape_2021,wang_transient_2022,guan_migration_2024}, and is equivalent to the small wave-number expansion in the complex Fourier space \citep{peng_upstream_2020,peng_rotational_2024}.
Here, due to the inherent unsteadiness of the forcing flow, a two-time-scale homogenisation is needed.

We first rewrite \cref{eq:transport equation dimensionless} in a co-moving streamwise coordinate, $x_0 \triangleq  x - \overline{U_d^{\infty}} t$, where $\overline{U_d^{\infty}}$ is the period-averaged drift:
\begin{equation}
	\frac{\partial P}{\partial t}+ \left(U_x - \overline{U_d^{\infty}}\right) \frac{\partial P}{\partial x_0} - D_t \frac{\partial^2 P}{\partial x_0^2} + \mathcal{L}_{co} P = 0.
	\label{eq:comoving eq}
\end{equation}
Note that while we use the same symbol $\overline{U_d^{\infty}}$ here, its value is yet to be determined from the perturbation equations, distinct from its previous use in \cref{sec:GTD}.

We then consider the two time scales:
\begin{equation}
	\tau_0 \triangleq t, \quad \tau_1 \triangleq \epsilon^2 t,
\end{equation}
where $\tau_0$ represents the fast time scale compared with the slow diffusive time scale $\tau_1$, since $\epsilon$ is a small parameter.
\Cref{eq:comoving eq} is rewritten as:
\begin{equation}
	\frac{\partial \tilde{P}}{\partial \tau_0} + \epsilon^2 \frac{\partial \tilde{P}}{\partial \tau_1} + \left(U_x - \overline{U_d^{\infty}}\right) \frac{\partial \tildeP}{\partial x} - D_t \frac{\partial^2 \tildeP}{\partial x^2} + \mathcal{L}_{co} \tildeP = 0,
\end{equation}
where $\tildeP(x,z,\theta,\tau_0,\tau_1) = P(x,z,\theta,t)$.

We next introduce the typical diffusive length scaling:
\begin{equation}
	\xi = \epsilon x_0,
\end{equation}
here $\epsilon$ is a small parameter, which can be interpreted as $W^{\ast}/L^{\ast}$, with $L^{\ast}$ denoting the characteristic streamwise length scale of the dispersing swimmer patch.
Under this rescaling, the transport equation becomes:
\begin{equation}
	\frac{\partial \tilde{P}}{\partial \tau_0} + \epsilon^2 \frac{\partial \tilde{P}}{\partial \tau_1} + \epsilon \left(U_x - \overline{U_d^{\infty}}\right) \frac{\partial \tildeP}{\partial \xi} - \epsilon^2 D_t \frac{\partial^2 \tildeP}{\partial \xi^2} + \mathcal{L}_{co} \tildeP = 0.
	\label{eq:governing equation tilde P}
\end{equation}

We expand $\tildeP$ as a regular perturbation series in powers of $\epsilon$:
\begin{equation}
	\tildeP = \tildeP_0 + \epsilon \tildeP_1 + \epsilon^2 \tildeP_2 + O(\epsilon^3).
	\label{eq:perturbation order}
\end{equation}
To ensure solvability at each order, we impose the following solvability conditions:
\begin{subequations}
\begin{equation}
	\int_0^1 \int_0^{2\upi} \tildeP_1 \, \mathrm{d}\theta \, \mathrm{d}z=0,
\end{equation}
\begin{equation}
	\int_0^1 \int_0^{2\upi} \tildeP_2 \, \mathrm{d}\theta \, \mathrm{d}z=0,
\end{equation}
\end{subequations}
together with periodicity conditions on $P_1$ and $P_2$ in $\tau_0$, each with a period of $T$.

The perturbation equations at successive orders of $\epsilon$ are obtained by substituting \cref{eq:perturbation order} into \cref{eq:governing equation tilde P} and collecting terms of the same order:
\begin{subequations}
\begin{equation}
	O(1): \frac{\partial \tildeP_0}{\partial \tau_0} + \mathcal{L}_{co} \tildeP_0  = 0, 
	\label{eq:perturbation O1}
\end{equation}
\begin{equation}
	O(\epsilon): \frac{\partial \tildeP_1}{\partial \tau_0} + \left(U_x - \overline{U_d^{\infty}} \right) \frac{\partial \tildeP_0}{\partial \xi} + \mathcal{L}_{co} \tildeP_1 = 0, 
	\label{eq:perturbation O epsilon}
\end{equation}
\begin{equation}
	O(\epsilon^2): \frac{\partial \tildeP_2}{\partial \tau_0} + \frac{\partial \tildeP_0}{\partial \tau_1}  + \left(U_x - \overline{U_d^{\infty}}\right) \frac{\partial \tildeP_1}{\partial \xi} - D_t \frac{\partial^2 \tildeP_0}{\partial \xi^2} + \mathcal{L}_{co} \tildeP_2 = 0.
	\label{eq:perturbation O epsilon square}
\end{equation}
\end{subequations}

The leading-order solution $\tildeP_0$ is assumed to take the separable form
\begin{equation}
	\tildeP_0(\xi, z, \theta, \tau_0, \tau_1) = g_0(z, \theta, \tau_0) C(\xi, \tau_1),
\end{equation}
where $g_0$ represents the conditional probability density in the $(z,\theta,\tau_0)$ space, and $C$ denotes the streamwise, period- and cross-section-averaged concentration.
Substituting this ansatz into the $O(1)$ perturbation problem yields the governing equation for $g_0$:
\begin{equation}
	\frac{\partial g_0}{\partial \tau_0} + \mathcal{L}_{co} g_0 = 0,
	\label{eq:governing equation g0}
\end{equation}
with periodic conditions $g_0|_{\tau_0} = g_0|_{\tau_0+T}$.
\vskip 5pt

The first-order correction $\tildeP_1$ is assumed to take the form
\begin{equation}
	\tildeP_1(\xi,z,\theta,\tau_0,\tau_1) = \chi(z,\theta,\tau_0) \frac{\partial C}{\partial \xi} + g_0(z,\theta,\tau_0) h(\xi,\tau_1),
\end{equation}
where $\chi$ satisfies the same boundary and periodic conditions as $g_0$.
Substituting this expression into the $O(\epsilon)$ perturbation equation yields the cell problem for $\chi$:
\begin{equation}
	\frac{\partial \chi}{\partial \tau_0} + \mathcal{L}_{co} \chi = \left(\overline{U_d^{\infty}} - U_x\right) g_0 .
	\label{eq:governing equation chi}
\end{equation}
Integrating the above equation over $\tau_0 \in[\tau_0, \tau_0+T]$, $z\in[0,1]$, and $\theta \in[0,2\upi]$ eliminates the left-hand side by boundary and periodic conditions, leading to the expression for the period-averaged drift:
\begin{equation}
	\overline{U_d^{\infty}} = \frac{1}{T}\int_{\tau_0}^{\tau_0+T} \int_0^1 \int_0^{2\upi} U_x g_0\, \mathrm{d} \theta \, \mathrm{d} z \, \mathrm{d} \tau_0.
	\label{eq:expression Ud homogenisation}
\end{equation}

Applying the solvability condition, i.e., integrating the $O(\epsilon^2)$ 
perturbation equation \cref{eq:perturbation O epsilon square} over $\tau_0 \in[\tau_0, \tau_0+T]$, $z\in[0,1]$, and $\theta\in[0,2\upi]$ yields the effective one-dimensional dispersion equation:
\begin{equation}
	\frac{\partial C}{\partial \tau_1} =  \overline{D_T^{\infty}} \frac{\partial^2 C}{\partial \xi^2},
	\label{eq:expression DT homogenisation}
\end{equation}
where $\overline{D_T^{\infty}}$ denotes the period-averaged effective dispersivity with the same symbol with \cref{sec:GTD}.
The expression for $\overline{D_T^{\infty}}$ is given by
\begin{equation}
	\overline{D_T^{\infty}} = D_t + \frac{1}{T}\int_{\tau_0}^{\tau_0+T} \int_0^1 \int_0^{2\upi} \left[ (\overline{U_d^{\infty}} - U_x) \chi \right] \, \mathrm{d} \theta \, \mathrm{d} z \, \mathrm{d} \tau_0.
\end{equation}
Comparing the governing equations and resulting expressions \cref{eq:governing equation g0,eq:governing equation chi,eq:expression Ud homogenisation,eq:expression DT homogenisation} with their counterparts in the generalised Taylor dispersion framework, specifically \cref{eq:governing equation asymptotic P0,eq:governing equation bN,eq:expression Ud mean GTD,eq:expression DT mean GTD}, it becomes evident that the two-time-scale homogenisation method is formally equivalent to the GTD theory in capturing the long-time asymptotic period-averaged drift and dispersivity.
However, we note that the current two-time-scale homogenisation method cannot capture the temporal variations of drift and dispersivity within a period.
Such variations, which are retained in our extended GTD formulation, are averaged out in the homogenisation framework.

It is also important to distinguish the current two-time-scale homogenisation from the two-time-variable expansion introduced in \cref{sec:transient solutions}.
In the two-time-variable expansion, $t_0$ and $t_1$ do not necessarily have different orders of magnitude, no perturbation analysis is required, and the solutions for the transient moments equations are exact.
In contrast, the two-time-scale homogenisation method aims to capture the long-time asymptotic period-averaged dispersion behaviour by introducing two time scales of different order: the normal time scale $\tau_0$ and the asymptotic effective diffusive time scale $\tau_1$, given that $\epsilon$ is a small parameter.

Converting back to the original coordinate, \cref{eq:expression DT homogenisation} becomes the period-averaged one-dimensional dispersion equation:
\begin{equation}
	\frac{\partial C}{\partial t} =  \overline{U_d^{\infty}} \frac{\partial C}{\partial x} + \overline{D_T^{\infty}} \frac{\partial^2 C}{\partial x^2},
	\label{eq:effective 1D dispersion equation}
\end{equation}
which is valid in the long-time asymptotic limit.

\end{appen}

\bibliographystyle{jfm}

\end{document}